\documentclass[preprint]{aastex}
\newcommand{\Ha}{H$\alpha$ }
\newcommand{\Hza}{H$\alpha$}

\newcommand{\HeII}{He~II~$\lambda$4686 }
\newcommand{\OIII}{[O~III]~$\lambda$5007 }
\newcommand{\OzIII}{[O~III]~$\lambda$5007}

\newcommand{\NII}{[N~II]~$\lambda$6584 }
\newcommand{\NzII}{[N~II]~$\lambda$6584}

\newcommand{\HST}{{\it Hubble Space Telescope }}
\newcommand{\HzST}{{\it Hubble Space Telescope}}
\newcommand{\etal}{et~al.}

\newcommand{\VisAst}{
\noindent Visiting Astronomer, Kitt Peak National Observatory, National Optical
Astronomy Observatories, which is operated by the Association of Universities
for Research in Astronomy, Inc.\ under contract with the National Science
Foundation.\par} 
\newcommand{\aca}{Acta~Astron.}
\newcommand{\astk}{Astron.~Tsirk.}
\newcommand{\ha}{Harvard~Ann.}
\newcommand{\hb}{Harvard~Bull.}
\newcommand{\ibvs}{IBVS}

\newcommand{\jo}{J.~Obs.}

\newcommand{\obs}{Observatory}

\newcommand{\perz}{Perem.~Zvezdy}
\begin{document}

\title{Optical Imaging of Nova Shells and the Maximum Magnitude-Rate
of Decline Relationship}

\author{Ronald A. Downes\altaffilmark{1}}
\affil{Space Telescope Science Institute, Baltimore, MD 21218}
\email{downes@stsci.edu}

\and

\author{Hilmar W. Duerbeck\altaffilmark{2}}
\affil{WE/OBSS, Free University Brussels (VUB), Pleinlaan 2, 1050 Brussels,
  Belgium and Space Telescope Science Institute, Baltimore, MD 21218}
\email{hduerbec@vub.ac.be}

\altaffiltext{1}{\VisAst}
\altaffiltext{2}{based on observations collected at the European Southern
  Observatory, La Silla, Chile}

\begin{abstract}

An optical imaging study of recent 30 novae has been undertaken using
both ground-based and space-based observations.  Resolved shells have
been detected around 9 objects in the ground-based data, while another
four objects have shells detected by \HST observations; for RW~UMi, we
fail to detect a shell which was observed five years earlier.  Images
in \Hza, and when appropriate \OzIII, are shown, and finding charts
for novae without shells are given if no published chart is available.
Expansion parallaxes for all systems with shells are derived, and
absolute magnitudes for a total of 28 objects are presented, along
with a discussion of the maximum magnitude-rate of decline
relation. We find that separate linear fits for fast and slow novae
may be a better representation of the data than a single, global
fit. At minimum, most novae have similar magnitudes as those of dwarf
novae at maximum and novalike stars.

\end{abstract}


\section{Introduction}

In the course of a nova eruption, a significant amount of material
($10^{-4}$~M$_{\sun}$) is ejected, which can form a visible shell
around the system.  Observation of this shell can lead to accurate
distances to the novae via the expansion parallax method.  For this
reason, searches for nova shells have been undertaken by various
groups (e.g. \citet{cr83}, \citet{sod95}).  With the distance in hand,
one can then determine the luminosity of the system at maximum light,
which can be used to calibrate the maximum magnitude-rate of decline
relation.  Use of this relation can allow for determinations of the
distance to objects such as the Virgo Cluster \citep{dvl95}.

When we embarked on a survey of nova remnants with small
telescopes to measure nova shell line fluxes, we did not expect to
make a contribution to the study of nova shell morphology and the
nova luminosity calibration.  Nevertheless, the results (including the
discovery of 4 previously undetected shells), combined with data
collected at larger earthbound telescopes and with \HzST, encouraged us to
attempt a derivation of nova distances and luminosities at maximum 
and minimum. We constructed maximum magnitude-rate of decline relations, 
based on the standard light curve parameters $t_2$, $t_3$ and $m_{15}$,
and also studied the distribution of minimum magnitudes.
 
The use of novae as distance indicators poses more problems than that of
other methods based on variable stars. While novae are brighter than
pulsating variables, their light curve characteristics require a
monitoring with high time resolution (a day or better), and the 
determination of the decline rate requires close monitoring until 
a given object declines by two or three magnitudes. 

The nova luminosity calibration might be done on a purely photometric
basis, using (unreddened) novae in extragalactic systems (M31,
Magellanic Clouds).  However, we would need to tie the novae into the
distance ladder by pulsating variables or binaries, thus introducing
possible systematic errors of the distance scale.

The other approach is the
expansion parallax method, which combines the angular expansion rate 
(observed several years or decades after outburst) with radial velocity
observations during outburst or at later stages. This approach is also
open to systematic errors. Well-observed novae give evidence that mass ejection
is not spherical, and inclination effects play an important role. 
Furthermore, the radial velocity of the shell of a given nova is often
variable with time, even if we restrict ourselves to the observation of the
``principal system''. High and low velocity material is present, the slower
material sometimes shows signs of acceleration. Similar problems arise in
the determination of angular expansion rates, since the bright nebular 
structures are often immersed in a more extended fuzz. The more rapidly
expanding shell becomes increasingly fainter (a phenomenon which is also seen
in the `sharpening' of the nebular emission line profiles in nova spectra
in the course of the outburst), making the measurement of the shell size
more difficult.
If we restrict ourselves to the spectral emission line maxima and brightest 
features in the resolved shells, we still encounter a major problem, since 
these condensations are often arranged in an inclined ``equatorial ring''.

A combination of both methods is the derivation of the form of the
Maximum Magnitude-Rate of Decline (MMRD) calibration from an
extragalactic sample, while determining its zero point calibration
using galactic novae.

Additional problems are the determination of the proper light maximum
(sometimes it is uncertain whether the observed maximum is the true maximum,
especially when spectral observations in early stages are lacking) and the
amount of interstellar absorption (most galactic novae are concentrated
towards the galactic plane).

Galactic novae with a well-covered maximum (to obtain the rate of
decline), UV observations at the nebular or minimum stage (to derive
the reddening from line ratios or the 2200~\AA{} band), and a
well-resolved shell (to derive the kinematic characteristics), are
extremely rare. Only a handful of novae fulfill these criteria, and a
luminosity calibration based on small number statistics may suffer
from systematic effects, such as the effect of white dwarf masses or
chemical composition on the luminosity.  Thus we tried to use all
novae for which expansion parallax data are available, by making at
some points ``educated guesses'' or using ``brute averaging''. This
certainly puts a lot of arbitrariness in the calibration, but we hope
that this fairly extensive data base, which will not be augmented
considerably for a long time, gives a reasonable insight in the
present state of the art of novae as distance calibrators, if one
deals with available, and not ideal, samples.

\section{Observations}

The observations were obtained on 1998~March~21-23 with Dutch 0.9m
telescope at the ESO La Silla Observatory, using a TEK TK512CB chip ($512
\times 512$ pixels) with a scale of $0\farcs 465$/pixel.
Further observations were obtained at the Kitt Peak National
Observatory on 1998~May~28 - June~1 with the 2.1m telescope using the
TEK ``T1KA'' chip ($1024 \times 1024$ pixels with a scale of
$0\farcs 305$/pixel) and on 1998~June~30 with the 0.9m telescope
using the TEK ``T2KA'' chip ($2048 \times 2048$ pixels with a 
scale of $0\farcs 7$/pixel).  The data were obtained with narrow band
filters centered at \Ha (80\AA~FWHM at ESO, 36\AA~at KPNO) and \OIII
(55\AA~at ESO, 31\AA~at KPNO), as well as off-band and {\it UBVR}
filters; see Table~\ref{tbl-1} for details.  The data were reduced in the
standard manner.

The search for shells was performed using two different techniques.
First, a PSF was generated for each image (with the routines in
DAOPHOT) using, whenever possible, stars of comparable brightness to
the nova.  The PSF was then subtracted from each star in the image,
and the resultant image examined for evidence of extended emission
(i.e. residual flux) around both the nova and nearby stars (as a
control sample). In almost every case, this technique was definitive.
The second method was to perform a radial profile fit of the nova and
several stars of comparable magnitude.  The fits were compared to
determine if the nova had a broader profile.  When a shell was
detected, a contour plot was generated from which the size of the
shell was measured by determining the edge of the shell (i.e. where the
shell merged into the sky).

For shell sizes obtained from the literature, we note in Table~\ref{tbl-2} 
the measurement technique (direct image or radial profile) used. Radial
profiles were quoted for only four objects.  In two cases (RW~UMi and
V351~Pup), the two techniques yielded consistent answers, while in the 
remaining two cases (PW~Vul and QU~Vul), the radial profile values were
considerably smaller than the direct image values.  To maintain as
consistent a dataset as possible, these discrepant radial 
profile values were not used
in our analysis.

\section{Results}
\subsection{General remarks}

We have detected shells around 10 objects, and in Table~\ref{tbl-2} we
present the measured diameters from our data, as well as summarize the
previous measurements from the literature.  We have obtained data on
another 5 objects in which we fail to detect a shell, but for which a
shell has been reported in the literature or can be seen in WFPC2
data.  Since different measurement techniques can produce different
results, we include in the table a code indicating the measurement
technique used (radial profile fit or direct image); $85\%$ of the
measurements are from direct images where the shell is clearly
resolved.  For the radial profile fits, a correction for the PSF has
been made, while for the direct images, not such correction was made.
We have not corrected our measurements of direct images for geometric 
effects caused by the ellipticity of the shell (see below).  

In the following sections 4 and 5, we present the results of our new
observations (shell sizes, expansion rates), and derive distances and
luminosities; we include the supporting data necessary for these
derivations (i.e. interstellar extinction, light curve parameters).

Nebular expansion parallaxes are a fundamental method for determining
the distances to novae if the three basic parameters --- age,
expansion velocity, and size --- can be properly determined.  While age can
(usually) be determined accurately, size (geometry) and expansion velocity
are less certain, and will dominate the uncertainty in the derived distance.  
To derive a luminosity furthermore requires an accurate light curve and
a trustworthy reddening estimate.

\subsubsection{Geometry and kinematics}

Projected nova shells sometimes appear 
spherically symmetric, but in many cases are oval. Furthermore, the
densest gas clouds are often concentrated in a ring, which is inclined
to the line of sight of the observer. Thus, projection effects should
be taken into account (see \citet{whc00}). There is general agreement that 
if nova shells are non-spherical, they are prolate ellipsoids. 
From the shells of the slow novae T Aur, DQ Her, BT Mon, RR Pic, and 
FH Ser, whose inclinations can be estimated with some confidence, 
an average ellipticity $0.75\pm 0.05$ is deduced; if major diameters, or some
average diameters are used (as in our diameter and radial profile data), 
errors should on the average be within 10\% of the true distance. Other 
novae like GK Per or V603 Aql tend to have more spherical shells, or at 
least `condensations' arranged in a circular pattern.

Radial velocities show a very large
dispersion, not only in the different ``spectral systems'' (see,
e.g. \citet{mcl60}), but also for a given system. If a P Cygni line
profile is observed soon after outburst, the radial velocity of the
absorption line is strongly influenced by the superimposed emission,
but such radial velocities are the only information available
for some objects. In the nebular stage, ``castellated'' emission line
profiles are observed. They have broad wings that generally disappear
at later phases. Direct images of nova shells often show the associated
condensations (rings or blobs), surrounded by a faint fuzz, which
originates in faster moving material (as seen, e.g, deep images of nova shells
in \citet{due87c}). For several objects in our survey,
bright condensations are still close to the star and are easily resolved on
HST images, while more extended faint fuzz shows up on images taken with
earthbound telescopes. Ideally, one should be able to associate maxima of
emission lines with given condensations in the shell, i.e. one should have a
clear idea about the geometry of the shell. However, the
clumps most easily seen on direct images are those moving at right
angles to the line of sight, and do not contribute to the width or
``castellation'' of the emission line. In most cases, spatial
information of shell spectra, which is essential for reconstructing the
shell geometry, is missing. 

If there was no deceleration of the expanding shell, then measurements 
at any age would suffice for an accurate distance measurement.  However,
depending on the neighboring environment, nova shells could suffer
from deceleration. \citet{due87a} presented a method to approximate
the temporal variation of the shell radius, which can be applied if
observations at many epochs are available.  As only HR~Del and V603~Aql
have sufficient data for a deceleration determination, we have derived
distances to all objects assuming no deceleration. Our estimates 
are given in Table~\ref{tbl-3}.

\subsubsection{Luminosity}

The information needed to determine the 
absolute magnitude at maximum $M_{\rm max}$ or 15 days after maximum
$M_{15}$, which is essential for the use of a nova as a ``standard candle'',
is:

1. A trustworthy apparent magnitude at maximum (either a photoelectric 
$B$, $V$, visual ($v$) or photographic (pg) magnitude), as well as a 
well-defined light curve. In all cases, 
original sources have been re-examined, and light curve parameters have been
re-determined, rather than taken from data collections like \citet{due87b}.
We prefer photoelectric $V$ magnitudes over visual $v$
magnitudes, and only quote photographic (pg) magnitudes if no other
ones are available. 

2. The value of the interstellar reddening can be derived with sufficient 
accuracy from multi-wavelength observations of objects in decline 
(e.g. from emission line ratios or the strength of the 2200~\AA~ band), or from
UV observations of old novae (strength of the 2200~\AA~band). In cases
where no such information was available, $A_V$ was determined by the 
help of the galactic extinction program of \citet{hak97}, which interpolates 
and averages data based on various studies of galactic extinction.

In addition, the determination of the absolute magnitude of postnovae requires
a reliable magnitude in the minimum stage. We took $V$-magnitudes from our
survey, as well as from \citet{szk94} and a forthcoming statistical study 
of postnova luminosities \citep{dsls00}. 

In the following, absolute magnitudes of novae at maximum light are usually
given only with an accuracy of $0\fm 1$, and at minimum with an accuracy of
$0\fm 05$. Calculations were often done with higher accuracy, and
when corrections (photographic to visual magnitudes, effects of accretion disc
inclination) were applied, the result was rounded to the same accuracy.

\subsection{Individual objects}

\subsubsection{CP Pup (1942)}

CP~Pup was a fast nova that reached a maximum of $m_{v,\rm max} =
0.35$ in November~1942. Visual and photographic light curves were
published by \citet{pet49} and \citet{gap46}, from which we derive
$t_2$ = 6, $t_3=8$ days (averaged over the visual and photographic curves),
and $m_{v,15}= 4.95$, where $t_2$ is the time for a nova to decline 2
magnitudes from maximum, $t_3$ is the time to decline by 3 magnitudes,
and $m_{15}$ is the magnitude 15 days after maximum.  

The first reported detection of a shell was by
Zwicky (see \citet{bow56}), who noted a diameter of $5\farcs 6$ in
1955 in red light.  \citet{wil82} observed the shell in 1980
(\Hza/[N~II]) and found a diameter of 14\arcsec, while \citet{cr83}
found the same shell size (\Hza/[N~II]) in 1981. More recently,
\citet{go98} found a diameter of 20\arcsec~(\Hza/[N~II]) in 1995.  A
\HST image, obtained with the WFPC2 and F656N (\Hza) filter in 1995
(program 6060), shows the stellar remnant surrounded by a circle of
bright knots of diameter $19\farcs 8$; the average surface brightness
in the shell is $1.66 \times
10^{-15}\rm ~ergs~cm^{-2}~s^{-1}~arcsec^{-2}$.  Our \Ha image
(Figure~\ref{fig1}) reveals about 11 blobs arranged in a circular shell $\sim
20$\arcsec~in diameter (average surface brightness is $3.41 \times
10^{-14}\rm ~ergs~cm^{-2}~s^{-1}~arcsec^{-2}$). The same arrangement is seen 
in our [O~III] data (average surface brightness of $3.37
\times 10^{-14}\rm ~ergs~cm^{-2}~s^{-1}~arcsec^{-2}$).
The clumpiness seen in the 1980 and 1995 frames is still present.

The expansion velocity ($v_{\rm exp}$) for CP~Pup has been determined
by \citet{mcl43} and \citet{san45}, who derived a value of 1500~km/s
shortly after outburst. Spectra in the nebular stage \citep{san45}
show a profile consisting of 11 emission components, with a velocity
spread between $-454$ and +570~km/s. Similar results were obtained by
\citet{gra53} ($-510$ and +590~km/s). Spectroscopy of the resolved
blobs in the shell seen in the ground-based images yields a similar
range in velocities \citep{sei90}. We have assumed a $v_{\rm exp} =
550$ km/s, based on Gratton's result.

Previous distances estimates (\citet{due81}, \citet{wil82}, and
\citet{go98}) place CP~Pup at distances of 1.5, 1.6 and $1.8\pm
0.4$~kpc, respectively, while \citet{cr83} give 0.85~kpc, by neglecting the
shell geometry and assuming a much smaller expansion velocity. 

Since this is a well-observed object, we will illustrate
the uncertainties in the expansion rate that directly enters
into the distance determination. We compare the
ground-based deep \Ha frame, obtained with the ESO/MPIA 2.2\,m
telescope in 1987 \citep{due87c}, the
snapshot HST image of 1995, and our \Ha frame, obtained 1998 with the
ESO/Dutch 0.92\,m telescope. The expansion rate derived from the 
centers of the blobs that are arranged in the circular shell
is 0.151, 0.153 and 0.151 arcsec/year, respectively. In spite of 
the fact that the limiting magnitude and the seeing effects are 
very different for the three frames, the results are in good agreement.

If the size outlined by the outer edges of the blobs is used, the
expansion rates are 0.193, 0.187, and 0.178; furthermore, the 1987 
image reaches to fainter surface magnitudes, and the extent of the
``faint fuzz'' leads to an expansion rate of $0\farcs 26$/yr. 

The HST exposure reveals that instead of eleven blobs, the shell consists of
about 30 blobs with typical diameters of $1"$ (some are more extended and
show special shapes) The image also shows a symmetry axis that goes through 
several blobs which may be interpreted as ``polar blobs'', while the remainder
form an ring that is fragmented into $\sim$ 20 single clouds. If the ring is
circular, its ellipticity indicates an inclination of $\sim 42\fdg 5$,
and the blob-ring structure would form a prolate ellipsoid with $b/a=0.6$.

Correspondingly, the radial velocities observed in high-resolution spectra 
a few years after outburst, assumed to form principally in the fragments of
the ring, must be corrected for inclination. Thus, while the measured 
radial velocity is $\pm 550$ km/s, the true $v_{exp} = 550/\sin
42\fdg 5\sim 815$ km/s, and the distance, derived form the expansion 
rate of the blob centers (along the major axis), is 1140 pc. If this 
distance and the expansion rate of the faint fuzz are used to calculate
the radial velocity of the ``fast material'', a value of 1415 km/s is derived,
which is similar to that observed in the absorption lines in the spectrum of
the nova observed shortly after maximum. Such a dichotomy of fast and slow
moving material, forming shells of different diameters, also occurs in 
PW Vul, QU Vul, V842 Cen and V1974 Cyg (see below).

\citet{gil94} derived an $E_{B-V}$ value of $0.25$. 
Thus, the absolute magnitude at maximum is
$M_V=-10.7^{+0.6}_{-0.4}$, $M_{V,15}=-6.1$, and the absolute magnitude at
minimum is $M_V=3.90$.

Summing up, if the true expansion velocity is known, 
the use of ``outer edges'' would decrease the distance by 20\%,
the neglect of the inclination angle would decrease the distance by 32\%,
and a ``wrong'' expansion velocity of 1400-1500 km/s would increase the
distance by $\sim 80\%$. If these errors are taken to be typical ones, 
luminosities of objects for which details on 
shell structure and extent, inclination and radial velocity are not known, 
may suffer errors of up to $\pm 1\fm 25$. 

\subsubsection{CT Ser (1948)}

CT~Ser was discovered well past maximum light in early 1948, and was
monitored intensely during decline. The nova was observed again by
\citet{sh92}, who report a typical CV spectrum containing Balmer and
\HeII emission.  There have been no reports of a detection of a shell for 
this object.  Our \Ha image, shown in Figure~\ref{fig2}, clearly reveals 
a nearly circular shell in the PSF-subtracted image with a diameter of 
$\sim 8$\arcsec~in both directions (with an average surface brightness 
of $7.48 \times 10^{-16}\rm ~ergs~cm^{-2}~s^{-1}~arcsec^{-2}$). The shell 
does not appear in the [O~III] data (with a $3\sigma$ limiting surface 
brightness of $1.13 \times 10^{-15}\rm ~ergs~cm^{-2}~s^{-1}~arcsec^{-2}$).

Although they do not report on the existence of a shell, \citet{cr83}
do provide $v_{\rm exp} = 535$~km/s for CT Ser.

There are no previous distances estimates for CT~Ser.  Based on the
above expansion velocity, we derive a distance of 1.4~kpc.
According to the galactic extinction model of \citet{hak97}, the
extinction is low ($A_V=0.02\pm 0.17$), and is neglected here.

The observed apparent maximum magnitude, $m_V=7.9$, is likely $3^{\rm
m}$ below the true maximum.  The observed absolute magnitude at maximum,
$M_V=-3.0$, supports the assumption that CT Ser was a slow nova 
discovered long after maximum. The absolute magnitude at minimum is 
$M_V=5.40$.

\subsubsection{RW UMi (1956)}

RW~UMi was a slow nova at high galactic latitude with a large outburst
amplitude. The photographic light curve by \citet{sat63} yields
$m_{\rm pg} = 6.1$ at maximum, $t_2 = 48$ days, $t_3 = 88$ days, and
$m_{\rm pg, 15}=6.5$, although it is not certain that the true maximum was
observed, since there is a gap of 16 days before the first positive 
observation, and there is no spectroscopic data taken
during outburst. A marginally resolved shell was originally detected by
\citet{coh85} with a diameter of $2\farcs 0$.  More recently,
\citet{sod95} observed a diffuse shell with a diameter of $3\farcs 0$,
as well as two regions of nebulous emission unconnected to the shell.
Our \Ha image, shown in Figure~\ref{fig3}, fails to detect either the
diffuse shell nor the two unconnected regions seen by \citet{sod95};
the $3\sigma$ limiting surface brightness is $5.88 \times
10^{-16}\rm ~ergs~cm^{-2}~s^{-1}~arcsec^{-2}$.  The shell is also
not visible in our [O~III] data to a limiting surface brightness of
$7.95 \times 10^{-16}\rm ~ergs~cm^{-2}~s^{-1}~arcsec^{-2}$.  A WFPC2
\Ha image obtained about five months prior to our observations (for
details see \citet{rwo98} -- program 7386) also failed to detect the
shell to a $3\sigma$ limiting surface brightness of $6.37 \times
10^{-16}\rm ~ergs~cm^{-2}~s^{-1}~arcsec^{-2}$.  In the five years
between the observations, it appears the shell has faded below the
detection limit.

\citet{coh85} estimated $v_{\rm exp}=950$~km/s.
Previous distance estimates place RW~UMi at a distance of $\sim
5.0$~kpc (\citet{coh85} estimated 5.6 kpc, \citet{kc89} $>2.8-6.9$ kpc and
\citet{sod95} $5\pm 2$~kpc). We adopt the latter result.
A value of $A_{V}=0.29\pm 0.24$ is derived from the program of
\citet{hak97}.  With an apparent maximum magnitude $m_{\rm pg} =6.1$
(which may not correspond to true outburst maximum), the absolute
magnitude at maximum is $M_{\rm pg}=-7.8^{+1.1}_{-0.7}$ (error from distance),
$M_{\rm pg,15} = -7.3$, and the absolute magnitude at postoutburst 
minimum is $M_V=5.15$.

\subsubsection{HR Del (1967)}

HR~Del was an unusually slow nova with a well-observed outburst.  The
$V$ light curve compiled by \citet{due96} yields $m_{V, \rm max} =
3.76$, $t_2 = 172$ days, $t_3 = 230$ days, and $m_{V,15} =
5.08$. \citet{sod76} first obtained spatially resolved spectroscopy
of the nebular shell, followed by a more detailed study by \citet{solf83}.
\citet{koh81} obtained the first photograph
of the resolved shell in [O~III], finding a oval remnant
$3\farcs 7 \times 2\farcs 5$ in size in 1981. \citet{sod94} obtained CCD
imaging in both [O~III] and \Hza/[N~II] in 1992, which
revealed a ring-like structure of diameter $5\arcsec - 6\arcsec$ in
\Hza/[N~II], and a bipolar structure of similar diameter in [O~III].
The \Hza/[N~II] also revealed enhancements that lie at the same
positions as the [O~III] components, although whether this implies
emission from the same material, or a simple line-of-sight alignment,
is not clear.  In 1993, \citet{sod95} re-observed HR~Del, revealing
more extended emission ($11\farcs 5 \times 8\farcs 5$) than previously
seen.  A \HST image, obtained with the WFPC2 and F656N (\Hza) filter
(program 6770), does not show the enhancements seen by \citet{sod95},
and has a smaller extent ($8\farcs 7 \times 6\farcs 9$); the average
surface brightness is $1.85 \times 
10^{-14}\rm ~ergs~cm^{-2}~s^{-1}~arcsec^{-2}$.

Our images, shown in Figure~\ref{fig4}, are similar to that in
\citet{sod94}, with \Ha having a complete shell ($9\farcs 8\times
8\farcs 5$), and a surface brightness ranging from $6.43 \times
10^{-14}\rm ~ergs ~cm^{-2} ~s^{-1} ~arcsec^{-2}$ in the outer region to
$1.85 \times 10^{-13}\rm ~ergs~cm^{-2}~s^{-1}~arcsec^{-2}$ in the
inner region.  The [O~III] image shows a bi-polar structure, with an
average surface brightness of $1.59 \times
10^{-13}\rm ~ergs~cm^{-2}~s^{-1}~arcsec^{-2}$.  The enhancements
seen by \citet{sod94} in the \Hza/[N~II] image are not present in our
image, and the emission in the 2 lines appear to be growing at
slightly different rates (i.e. the major axis in [O~III] is $9\farcs
5$).

The expansion velocity for HR~Del was measured by \citet{mal75}
(470~km/s), \citet{koh81} (540~km/s), \citet{cr83} (520~km/s), and
\citet{solf83} (560~km/s).

Previous expansion parallax measurements (\citet{mal75},
\citet{koh81}, \citet{due81}, \citet{solf83}, \citet{cr83}, \citet{sod94}, and
\citet{sod95}) yield a distance for HR~Del of $940\pm 155$~pc.
Distances based on other techniques (e.g. interstellar lines, reddening, 
light curve decline) yield
comparable values of $835\pm 92$~pc \citep{drd77}.  Based on an expansion
velocity of 525~km/s (similar to that used by the papers cited above),
we derive a distance of 750~pc.  Note that we attempted to fit the
HR~Del data to derive the deceleration, but the best fit resulted in
an acceleration of the shell.  An acceleration is possible if there is
a fast wind from the stellar remnant, and P~Cygni profiles (indicative
of outflow) have been detected in the spectrum of HR~Del
\citep{kkw81}.

The expansion rates for HR~Del show significant variations, with the
measurement of \citet{sod95} being the most discrepant.  That
observation was obtained with a similar filter and exposure time to
our data, but with a telescope twice as large, and therefore goes
much fainter.  This can be seen by overlaying our data on the
\citet{sod95} \Ha image, which shows that our shell is slightly larger
than the main emission in their image, and that we fail to detect
their lowest ($2\sigma$) contour.

We adopt a distance of $0.76\pm 0.13$~kpc. \citet{ver87} derived an
$E_{B-V}$ value of $0.15\pm 0.03$, and with an apparent maximum
magnitude $m_V =3.76$, the absolute magnitude at maximum is $M_V=
-6.1\pm 0.4$, $M_{V,15}= -4.8\pm 0.4$, 
and the absolute magnitude at minimum is $M_V=2.30$.

\subsubsection{V3888 Sgr (1974)}

V3888~Sgr was discovered in late 1974 during its decline from 
observed maximum,
which was rapid ($t_3 \sim 10$ days according to \citet{lwv76}).
While some photometric \citep{vm77} and spectroscopic \citep{lwv76}
observations were obtained during decline, there have been no
published observations of the object in quiescence.  Our \Ha image,
shown in Figure~\ref{fig5}, shows a nearly complete shell of size
$5\farcs 2\times 4\farcs 6$ (with a surface brightness of $7.82 \times
10^{-16}\rm ~ergs~cm^{-2}~s^{-1}~arcsec^{-2}$); the object is too
faint to yield a reliable profile fit.

Radial velocity measurements made shortly after discovery gave an
average $v_{\rm exp}$ of $\sim 1300$~km/s \citep{lwv76}.

Previous estimates place V3888~Sgr at a distance of 3.5~kpc
\citep{vm77} to $\lesssim 6$~kpc \citep{lwv76}, based on photometric
and spectroscopic evidence.  Our derived
distance, based on a $v_{\rm exp}$ of 1300~km/s, is 2.5~kpc.
The galactic extinction program of \citet{hak97} yields a value 
$A_V=1.21\pm 0.60$, however, \citet{vm77} argue for a value
$A_V=3.2$.

The observed apparent maximum magnitude, $m_V=9$, is likely $2\fm 5$
below the true maximum, as indicated from the spectroscopic appearance 
of the nova. If $A_V=1.21$ is used, the observed absolute magnitude, 
$M_V=-4.2\pm 0.6$ would point at a slow nova with $M_V \sim -6.5$.
In this case, the absolute magnitude at minimum, $M_V=7.75$, appears to be 
unusually faint. If the second value, $A_V=3.2$ is used, the observed 
absolute magnitude, $M_V=-6.2$, yields an extrapolated maximum luminosity
which assigns V3888 Sgr to the group of fast novae (an assumption 
that also entered the derivation of the reddening value of \citet{vm77}).
The absolute magnitude at minimum, $M_V=5.75$, would be in better 
agreement with other novae (this result was also used in Table~\ref{tbl-6}). 
Thus, either the galactic extinction program 
underestimates the reddening value, or the exnova is indeed unusually 
faint (e.g. the exnova lies in the period gap). 

\subsubsection{NQ Vul (1976)}

NQ~Vul was a moderately fast nova that has been well-studied.  The
light curve is somewhat peculiar (see the compilation by
\citet{due81}); the delayed maximum of $m_V= 6.25$ (with a scatter of
0.25) is followed by a short, deep minimum, so that a derived
$t_2$-time of 2.5 days is of no significance: if one disregards this
fluctuation, $t_2 = 23$ days, $t_3 = 53$ days, and $m_{V,15} = 7.55$.
Observations by \citet{sod95}, obtained in 1993, revealed a circular
shell with an average diameter of $\sim 8\arcsec$, a knot at the
northern end, and an inner shell with a diameter of $3\arcsec$.  There
is also evidence of a bar-like feature extending across the stellar
remnant at a position angle of $\sim 140\arcdeg$.  A WFPC2 \Ha image
obtained in 1997 (program 7386) did not detect the shell to a
$3\sigma$ limiting surface brightness of $6.88 \times
10^{-16}\rm ~ergs~cm^{-2}~s^{-1}~arcsec^{-2}$.  However, due to its
extremely high resolution, WFPC2 is not efficient at detecting
extended, diffuse objects, so the lack of detection is not unexpected.
Our \Ha image, shown in Figure~\ref{fig6}, reveals an almost circular
shell of size $7\farcs 3\times 7\farcs 0$, with an average surface
brightness of $8.98 \times 10^{-15}\rm ~ergs~cm^{-2}~s^{-1}~arcsec^{-2}$.  
The northern extension seen by \citet{sod95} is detected, and while 
a bar-like structure is seen, comparison with field stars shows that 
it is an artifact of the PSF subtraction. The shell does not appear 
in the \OIII data to a $3\sigma$ limiting surface brightness of
$2.24 \times 10^{-15}\rm ~ergs~cm^{-2}~s^{-1}~arcsec^{-2}$.

The absorption lines of NQ~Vul were measured soon after outburst
(\citet{cs78}, \citet{you80}, and \citet{kw78}), and a wide range of
expansion velocities are given by these authors. \citet{kw78} noted
the conspicuous high-velocity systems; they derived for the principal
spectrum an average value $-950$, for the diffuse enhanced $-1500$,
and for the Orion spectrum $-1950$~km/s. The total half-width at zero
intensity of the strong \OIII and Balmer lines is 1150 and 900 km/s,
respectively.

One year after outburst, \citet{kw78} observed a four-peaked structure 
with components at $-554, -295, +45$ and $+537$~km/s, and \citet{cr83} 
derived a $v_{\rm exp}$ of 705~km/s 5-6 years after outburst.  As
usual, the absorption lines have higher velocities than the emission
lines, which may indicate a noticeable inclination of the principal
part of the shell. The expansion velocity of \citet{cr83} may be an
acceptable compromise between the absorption and emission velocities
measured by \citet{kw78}.

Previous distances estimates place NQ~Vul at a distance of 1.7 kpc
\citep{kw78}, 1.2~kpc \citep{due81} and $1.6\pm 0.8$~kpc \citep{sod95}. 
If we associate the inner shell of \citet{sod95} with the slowly expanding 
material observed by \citet{kw78} (average $v_{\rm exp}$ of the fast components 
= $\pm 545$ km/s), and the larger shell, as well as our observation, 
with the material that is evident in the total width of the nebular and 
Balmer lines ($\pm 1025$ km/s), we derive an average distance 
$1.16\pm 0.21$~kpc. The nova polarization \citep{kw78} indicates 1.8 kpc.

In the following, we adopt the distance of 1.16~kpc. A value of 
$A_{V}=1.38\pm 0.49$ was derived from \citet{hak97}, while \citet{kw78} 
find a value of 2.79 from interstellar polarization. We have used
$A_V = 2.1\pm 0.7$ in the following.  
The absolute magnitude at maximum is $M_V=-6.1\pm 0.8$, $M_{V,15} = -4.85$, 
and the absolute magnitude at minimum is $M_V=5.30$.

\subsubsection{PW Vul (1984)}

PW~Vul was a moderately fast nova \citep{and91} with a 
well-studied outburst. The light curve is very structured, so that the
derivation of reliable decline times is difficult, as noted by
\citet{rs95}; we adopt a maximum magnitude $m_V = 6.3$, $t_2 = 82$
days, $t_3 = 126$ days, and $m_{V,15} = 7.7$. A spectrum of the
object, obtained in 1991 by \citet{rnm96}, shows strong \Ha and \OIII
emission.  The shell of PW~Vul was first detected by \citet{rn96} in
1993, who found a diameter of $1\farcs 1$ by examining the radial
profile of the object.  Our \Ha image, shown in Figure~\ref{fig7},
reveals a clearly resolved, circular shell of size $4\farcs 0\times
3\farcs 7$, with an average surface brightness of $9.35 \times
10^{-15}\rm ~ergs~cm^{-2}~s^{-1}~arcsec^{-2}$.  The shell does not
appear in the [O~III] data to a $3\sigma$ limiting surface brightness
of $2.95 \times 10^{-15}\rm ~ergs~cm^{-2}~s^{-1}~arcsec^{-2}$.
However, a WFPC2 exposure (program 7386) obtained shortly after our
observations reveals a faint, diffuse shell of size $1\farcs 5\times
1\farcs 5$, considerably smaller than our ground-based observations,
with an average surface brightness of $7.78 \times
10^{-15}\rm ~ergs~cm^{-2}~s^{-1}~arcsec^{-2}$.  The WFPC2 data is
detecting emission ``hidden'' in the ground-based PSF, while the
ground-based data shows diffuse emission which WFPC2 cannot detect.

In order to investigate the large difference in shell sizes between
the \citet{rn96} data and our data, the former authors kindly provided
us with a copy of their data.  We were able to confirm their radial
profile measurement, however a contour plot reveals a shell of size
$4\farcs 0\times 3\farcs 7$.  While this measurement has not been
corrected for the intrinsic PSF, it does illustrate the difference in
determining the shell size from radial profiles (FWHM) and from
contour plots (FWZI). 

The expansion velocity was measured by \citet{kw86} shortly after 
outburst to be 1200~km/s.
\citet{rn96} measured the velocity 9 years after outburst from the structure
of the H$\alpha$/\NII profile, and found a much lower value of 
$470\pm 60$~km/s.  
We associate the shell found by HST with the
material expanding with 470 km/s, and the more extended shell with 
the ``fuzz'' that expands with 1200 km/s, and obtain a consistent distance of
$1.8\pm 0.05$~kpc.
Previous distances estimates (\citet{dgn84}, \citet{sai91},
\citet{ads91} and \citet{rn96}) place PW~Vul at a distance of 1.2,
1.5-3.0, 1.3, and 1.6~kpc. The first three determinations use 
techniques other than expansion parallax (e.g. 
interstellar reddening measurements), while \citet{rn96} adopted
a small expansion velocity, in combination with an underestimated shell size,
to derive a result which is, nevertheless, in agreement with the others.

We adopt a distance of $1.8\pm 0.05$~kpc, and an interstellar extinction 
$A_V=1.73\pm 0.32$ \citep{and91}. The absolute magnitude at maximum 
is $M_V=-6.7\pm 0.4$, $M_{V,15}= -5.3$, and the absolute magnitude at 
minimum is $M_V=4.50$.

\subsubsection{QU Vul (1984)}

QU~Vul was a moderately fast nova of the ONeMg group, which was well
studied from the X-ray to the radio.  From the visual light curve from
the AFOEV archive, we derive a maximum magnitude $m_V = 5.6$, decline
times $t_2=22$ and $t_3=49$ days, and $m_{V,15} = 6.75$.  Radio
observations by \citet{tshp87} first detected the shell about one year
after outburst, and yielded a diameter of $0\farcs 23\times 0\farcs
13$ at an age of 497 days.  The first reports of an optical shell were
presented by \citet{dgb97}, who detected a larger radial profile for
the nova than for field stars in \Hza, while a continuum image showed
no differences.  Those authors derived a diameter for the shell of
$1\farcs 90$ in 1994.  \citet{shi98} obtained infrared images of the
object in 1996, and found a shell diameter of $0\farcs 8$ at 
$2.2~\micron$, again by comparing the radial profile of the nova with field
stars.  Data was obtained in 1998 with the \HST using both the WFPC2
(program 7386) and NICMOS (for details see \citet{belle99} -- program
7381).  Our measurements of that data reveal shell sizes of $1\farcs
6\times 1\farcs 3$ at $2.2~\micron$ and $1\farcs 6\times 1\farcs 7$ (a
bright, filled-in ring with an average surface brightness of $9.20
\times 10^{-14}\rm ~ergs~cm^{-2}~s^{-1}~arcsec^{-2}$) at \Hza.  Our
\Ha image, shown in Figure~\ref{fig8}, reveals a clearly resolved,
circular shell of size $5\farcs 6$ and an average surface brightness
of $1.32 \times 10^{-14}\rm ~ergs~cm^{-2}~s^{-1}~arcsec^{-2}$.  The
shell may be weakly seen in the [O~III] data, with an average surface
brightness of $4.19 \times
10^{-15}\rm ~ergs~cm^{-2}~s^{-1}~arcsec^{-2}$; the radial profile
plot clearly shows the presence of the shell.  The gross discrepancies
in the expansion rates between our data (based on a resolved shell)
and previous data (based on radial profiles) may indicate an
over-correction for the intrinsic PSF (hence an underestimation of the
shell size) when determining shell sizes based on profiles. As noted
by \citet{whc00}, the measurement technique of \citet{shi98} is
adequate for demonstrating the existence of a shell, but it is not
appropriate for determining the shell size, as the technique
incorrectly assumes a gaussian profile for the shell, and fails to
account for light from the central star.  The large discrepancy
between the ground-based and HST data may be caused by the fact that
the first method traces extended, rapidly moving, low-surface
brightness material, while the second one traces dense, clumpy, slowly
moving material near the central object.

The expansion velocity was measured shortly after outburst to be
790~km/s \citep{ri87} and 1200~km/s \citep{tshp87}.  A study of the
profiles of the \OIII and \Ha emission lines in early July, 1985, by
\citet{and91} yielded FWHM values of 1700 and 1360~km/s,
respectively (his FWZI is $\sim 3000$~km/s, the width of the flat-topped 
section of the \OIII{} line 1140 km/s). 
\citet{dgb97} derived a FWZI value of $1190\pm 105$~km/s 
in 1994.

Previous distances estimates, based on expansion parallaxes, 
place QU~Vul at distances of 3.6 and 2.6~kpc, respectively, 
\citep{tshp87, dgb97},
while photometric and reddening techniques yield distances of 3.0~kpc
\citep{geh85}, 1.2~kpc \citep{ber85}, and 1.6~kpc \citep{ads94}.
Based on an expansion velocity of 1500~km/s (the fastest-moving material), 
we derive a distance, based on our observations only, of 1.5~kpc. 
If the shell-structure of the HST observations is associated with the
material that formed the flat-topped structure in the lines observed by 
\citet{and91}, a distance of 2.0 kpc is derived. Averaging these 
results leads to $d=1.75\pm 0.25$~kpc, in reasonable agreement with
estimates based on other methods ($d=1.9\pm 0.5$~kpc).

With a distance of 1.75~kpc, and interstellar extinction
$E_{B-V}=0.60$ \citep{and91}, the absolute magnitude at maximum is
$M_V=-7.5\pm 0.25$, $M_{V,15} = -6.35$, and the absolute magnitude at 
minimum is $M_V=5.20$.

\subsubsection{V1819 Cyg (1986)}

V1819~Cyg was a slow nova that has been little studied; see
\citet{whi89} for observations of the nova in the first two years
after outburst.  The light curve has been constructed by
\citet{whi89}, who give $m_V({\rm max})~=~8.5\pm 0.1$; their $(B-V)$
value appears to be not reliable.  They derived times $t_2=21-59$~days
and $t_3=87-107$~days, respectively, although other authors give different
values. We have evaluated the AFOEV visual light curve, and derive
$m_V = 8.7$ at maximum, $t_2 = 37$ and $t_3 = 89$ days, and $m_{V,15} =
9.9$.  While no ground-based imaging observations have been reported,
a WFPC2 observation (program 7386) in the
F658N ([N~II]$\lambda6583$) filter reveals a marginally resolved shell
with a diameter of $0\farcs 5$.  Our ground-based observations are
unable to resolve a shell of that size.  \citet{szk94} has published
a finding chart for the object, but our images (see Figure~\ref{fig9}) have
resolved the Szkody object into two stars, with the nova being the
southeastern object.

From low-resolution spectral scans, \citet{whi89} derived ``velocity
widths'' of $1200\pm 200$~km/s.  \citet{and89} made spectroscopic 
observations at moderate dispersion one month after maximum 
and observed wide emission lines with a ``half-width'' of 1400~km/s 
(which we interpret as ``full width at half maximum''), with the 
principal and diffuse-enhanced absorption components at $-650$ and 
$-1450$~km/s.

\citet{whi89} find a reddening $E_{B-V}=0.35\pm 0.15$, and an absolute
magnitude at maximum $M_V=-7.3\pm 0.4$ from various relationships
between maximum magnitude and speed of decline, and derive a distance
of $8.7^{+2.8}_{-2.1}$~kpc.  If we assume, in agreement with
\citet{and89}, a $v_{\rm exp}= 700$~km/s for the bulk of
material, we derive a distance of 7.4~kpc, in good agreement with
the above result. This yields an absolute magnitude $M_V = -6.8\pm
0.5$, $M_{V,15} = -5.6$, and the absolute magnitude at 
minimum is $M_V=4.90$, with the error based on the uncertainty of
the extinction only.  The much larger value $A_V = 2.35\pm 0.95$ 
from the galaxy extinction model of \citet{hak97} would increase 
the brightness by $1\fm 25$.

\subsubsection{V842 Cen (1986)}

V842~Cen was a moderately fast nova ($t_2 = 35$ days, $t_3 = 48$ days,
\citep{sek89,ads94}), which showed a dust-forming event.
Observations by \citet{go98} revealed a shell of diameter $1\farcs
6$~(\Hza/[N~II]) in 1995.  Our \Ha image, shown in Figure~\ref{fig10}, 
reveals an incomplete, circular shell of size $5\farcs 6\times 6\farcs 0$,
with an average surface brightness of
$1.11 \times 10^{-12}\rm ~ergs~cm^{-2}~s^{-1}~arcsec^{-2}$. The shell
is also seen in the [O~III] data with an average surface brightness of
$1.16 \times 10^{-13}\rm ~ergs~cm^{-2}~s^{-1}~arcsec^{-2}$.

The expansion velocity was measured by \citet{sek89} after recovery from 
the dust-forming event in May 1988. The H lines showed a saddle-shaped 
structure, with the two peaks at $-630$ and $+440$~km/s. The line profiles 
showed broad wings which extended to $\pm 2000$~km/s (HWZI). A
study of the emission lines of \OzIII, \NII and \Ha by \citet{and91}
over the years 1988--1991 yielded a `castellated' profile, whose
initially broad wings faded, while two maxima at about $-500$ and
$+550$~km/s sharpened over that time interval. A high-resolution spectrum
of \OzIII, obtained on 1988 July 17, is shown in \citet{sei90}.  Its
maxima are at $-590$ and $+430$. The expansion velocity of the dense
regions of the shell is thus $\sim 525$~km/s, while low-density material
with $v_{\rm exp} = 2000$~km/s is also present.

Based on an expansion velocity of 525~km/s, \citet{go98} derived a
distance of $1.3\pm 0.5$~kpc.  The size of the shell in our data is
inconsistent with the size in \citet{go98}, and our distance of
0.42~kpc falls way below their value. The reddening-based distances of
0.9~kpc \citep{sek89} and 1.0~kpc \citep{ads94} tend to favor the
larger distance.  If we assume that the value of \citet{go98} is
correct, the expansion velocity of the shell observed by us is 1600
km/s, in reasonable agreement with the velocity of the material that
caused the broad wings observed in the first years following the
outburst.

\citet{and91} gave a light curve with a maximum magnitude $m_V=4.6$,
and $E_{B-V}=0.55\pm 0.05$; $m_{V,15}$ is $\sim 5.5$.  
A somewhat revised expansion rate and velocity of the shell observed by
\citet{go98} leads to a distance of 1.15~kpc, and the absolute magnitude 
at maximum is $M_V=-7.4^{+1.2}_{-0.8}$ (using the error estimate of 
the above authors), and $M_{V,15} = -6.5$, with the same error bars.
The nova has not yet returned to minimum light.

\subsubsection{QV Vul (1987)}

QV~Vul was a moderately fast nova, which was studied spectroscopically
by \citet{ads94}.  The AFOEV visual light curve yields $m_V=7.0$ at
maximum, $t_2=50$ days, $t_3=53$ days, and $m_{V,15}=7.95$.  While no
ground-based imaging observations have been reported, data was
obtained in 1998 with the \HST using both the WFPC2 (program 7386) and
NICMOS (program 7381).  Our measurements of that data reveal shell
sizes of $1\farcs 4\times 1\farcs 4$ at $2.2~\micron$, and $0\farcs
8\times 0\farcs 9$ at [N~II]$\lambda6583$ (with an average surface
brightness of $1.15 \times 10^{-13}\rm ~ergs~cm^{-2}~s^{-1}~arcsec^{-2}$).  
We failed to detect a shell in our ground-based images (to $3\sigma$ 
limiting surface brightnesses of $8.69 \times
10^{-16}\rm ~ergs~cm^{-2}~s^{-1}~arcsec^{-2}$ at \Ha and $1.11 \times
10^{-15}\rm ~ergs~cm^{-2}~s^{-1}~arcsec^{-2}$ at [O~III]).  There is
no published finding chart of the quiescent object, so we show, in
Figure~\ref{fig9}, our \Ha on-band and off-band images, which clearly
identify the nova.

An analysis of the shapes of \OIII and \NzII, observed in 1989 by
\citet{and91}, yields expansion (HWHM) velocities of $\pm 500$ and $\pm
530$, respectively.
Based on an averaged HWHM velocity of the \OIII and \NII lines,
$515$~km/s, we derive a distance to QV Vul of 2.7~kpc.

\citet{and91} gave a value of $E_{B-V}=0.40$. For a distance of
2.7~kpc, the absolute magnitude at maximum is $M_V=-6.4$,
$M_{V,15}=-5.45$, and the absolute magnitude at 
minimum is $M_V=5.20$.

\subsubsection{V351 Pup (1991)}

V351~Pup was a moderately fast nova which reached a maximum of at
least $m_v=6.4$, with $t_2=10$ days, $t_3 = 26$ days, and $m_{v,15} =
8.8$, as based on the AAVSO visual light curve \citep{mat99}. Its
outburst was studied in detail as part of the Tololo Nova Survey
\citep{wil94}.  A marginal ground-based detection of a shell (diameter
$0\farcs 58$) was reported by \citet{obd96}.  Data was obtained in
1998 with the \HST using the WFPC2 (program 7386), and our measurement
of that data reveals a \Ha shell $1\farcs 3$ in diameter, and an
average surface brightness of $2.87 \times
10^{-14}~\rm ergs~cm^{-2}~s^{-1}~arcsec^{-2}$.  Our ground-based
observations, shown in Figure~\ref{fig11}, do not show any indication
of a shell in the PSF-subtracted images.  However, in the radial
profile, the \Ha data does show that the nova (FHWM = 4.37~pixels) is
broader than field stars ($3.49 \pm 0.23$~pixels), while the off-band
image does not show any broadening of the nova profile.

The expansion velocity quoted by \citet{obd96}, 2000~km/s, is based on
IUE spectra. A detailed analysis of the Mg~II~2800 line, based on
high-resolution IUE spectra, resulted in an expansion velocity of
1200~km/s \citep{ps95}.

The expansion rate is about $0\farcs 100\pm 0\farcs 003$/yr, based on
the ground-based and HST observations. If the expansion velocity of
the Mg~II~2800 line is assumed to be representative for the shell, a
distance of $2.7\pm 0.2$~kpc is derived; the higher velocities quoted
by \citet{obd96} would lead to a distance around 4~kpc.

Maximum light is poorly covered by observations. \citet{sps96} derived
a reddening $E_{B-V}=0.72\pm 0.10$ from recombination lines. The
distance of 2.7~kpc leads to $M_V=-8.0\pm 0.4$, while the distance of 4~kpc
\citep{obd96} results in $M_V=-8.9\pm 0.4$; the values of $M_{V,15}$ are
$-5.6$ and $-6.5$, respectively. We will accept the average values as 
uncertain, but reasonable, estimates. The nova has not yet returned to 
minimum light.

\subsubsection{V1974 Cyg (1992)}

V1974~Cyg was a moderately fast nova, which has been extensively
studied.  The AFOEV visual light curve yields $m_v=4.4$ at maximum,
$t_2=17$ days, $t_3=37$ days, and $m_{v,15}=6.0$.  The first report of
a visible shell was by \citet{par94}, who used the Faint Object Camera
(FOC) on \HST 467 days after outburst to detect a circular,
inhomogeneous ring of diameter $0\farcs 26$ in the F278M (Mg~II)
filter; this observation was further analyzed by \citet{plhk95} and
they found the shell to be $0\farcs 22\times 0\farcs 28$ in size.
Further FOC observations \citep{plhk95} at \OIII (F501N) were obtained
at 690 and 818 days after outburst, and revealed shells sizes of
$0\farcs 30\times 0\farcs 42$ and $0\farcs 36\times 0\farcs 50$,
respectively.  \citet{ros96} detected extended \Ha emission around the
object (3\arcmin-4\arcmin) in 1994-1995, which they attribute to
excitation of the interstellar medium by UV photons from the outburst.
Our \Ha image, shown in Figure~\ref{fig12}, reveals a complete,
circular shell of size $8\farcs 6\times 9\farcs 0$, with an average
surface brightness of $3.66 \times
10^{-14}\rm ~ergs~cm^{-2}~s^{-1}~arcsec^{-2}$. The shell is also
seen in the [O~III] data, with an average surface brightness
of $7.21 \times 10^{-15}\rm ~ergs~cm^{-2}~s^{-1}~arcsec^{-2}$.
Further \HST observations were obtained in 1998 using both the WFPC2
(program 7386) and NICMOS (program 7381).  Our measurements of that
data reveal shell sizes of $1\farcs 4\times 1\farcs 3$ at
$2.2~\micron$, and $1\farcs 8\times 1\farcs 7$ (a bright ring with an
average surface brightness of $5.68 \times
10^{-14}\rm ~ergs~cm^{-2}~s^{-1}~arcsec^{-2}$) at \Hza.

The expansion velocity for V1974~Cyg was determined by several authors
to be 1100~km/s \citep{cgp97}, 1500~km/s \citep{sho93},
830-895~km/s \citep{ckh95}, and 700-950~km/s \citep{ros96}.

Previous distances estimates, based on expansion parallaxes from early
radio \citep{hje94} and \HST observations (\citet{par94},
\citet{plhk95}), place V1974~Cyg at a distance of $\sim 2.6$~kpc.
These observations were re-analyzed by \citet{cgp97}, who revised the
distance estimates downward to $1.7-1.9$~kpc, using expansion
parallaxes based on HST and radio data, and making them more
compatible with a distance of 1.8~kpc derived from other techniques
(\citet{chu93}, \citet{ros96}).  Based on an expansion velocity of
1070~km/s (which is the average value of the expansion velocities
given in the previous paragraph), we derive a distance for the \HST
data of 2.0~kpc.  However, if we use our ground-based observations,
where we see a clearly resolved shell, we derive a distance of only
310~pc.

This discrepancy can be removed in the same way as in the case of V842 Cen
and other objects.
The ground-based observations record faint, fast-moving material, while the
HST frames record the bulk of dense clouds that also show up as maxima in the
emission lines. If we assume a distance of 1.8~kpc, the extended shell
corresponds to an expansion velocity of 5700~km/s. The maximum expansion 
velocity seen in early P Cygni lines of Mg II was 4500 km/s \citep{sho93}.
The experimentum crucis would be to carry out spatially resolved spectroscopy
of the extended optical shell to see the velocity splitting.

We will adopt a distance of $1.8\pm 0.1$~kpc, based on Chochol's detailed
study, and an $E_{B-V}=0.35$ \citep{plhk95}, so that $M_V=-8.0\pm 0.1$, 
and $M_{V,15}=-6.4$. The nova has not yet returned to minimum light.

\subsubsection{HY Lup (1993)}

HY~Lup was discovered as a star of magnitude $V\sim 8$ by
\citet{lil93}, and has been little studied since. A spectrum taken by
the discoverer indicated that the nova was already some days after
maximum.  Photometry from the AAVSO archives \citep{mat99} reveals
$t_3 > 25$ days.  While no ground-based observations have been
reported since shortly after outburst, a WFPC2 observation (program
7386) in the F656N (\Hza) filter reveals a marginally resolved shell
with a diameter of $0\farcs 5$.  Our ground-based observations are
unable to resolve a shell of that size.  There is no published finding
chart of the quiescent object, so we show, in Figure~\ref{fig9}, our
\Ha on-band and off-band images, which clearly identify the nova.

The expansion velocity for HY~Lup was measured immediately after
discovery by \citet{del93} to be 2700~km/s. A spectrum taken three
years after outburst shows the object in the nebular stage; the
emission lines of H$\beta$, [O I] and [O III] appear double-peaked, with a
separation of $\sim \pm 800$~km/s and a FWHM of 340~km/s of the single
components \citep{due00}.

There are no previous distances estimates for HY~Lup.  Based on an
expansion velocity of 400~km/s, we derive a distance of 1.8~kpc.  This
estimate is based solely on \HST data, under the assumption that it
shows slow-moving material. If the \HST data also images fainter,
faster material, a larger velocity (up to 2700~km/s) would result in a
much larger distance.

The galactic extinction model yields a value $A_V = 0.7\pm 0.28$. The
resulting absolute magnitude is $M_V=-4.0$, but the expansion velocity
and the apparent magnitude may both be lower limits, so that the
absolute magnitude is likely some magnitudes brighter. The nova has 
not yet returned to minimum light. It cannot be  used for maximum
or minimum luminosity determinations, unless additional observations become
available.

\subsubsection{CP Cru (1996)}

CP~Cru was a fast nova ($t_2 \sim 4$ days, based on data from VSNET),
which showed strong He~II emission \citep{dvb96}, making it a member
of the He/N class \citep{wil92}. Its maximum observed magnitude was
$m_V = 9.25$, but true maximum may easily have been missed.  While no
ground-based imaging observations have been reported, data was
obtained in 1998 with the \HST using the WFPC2 (program 7386), and our
measurement of that data reveals a \Ha shell with a size 
$0\farcs 6 \times 0\farcs 6$. Our ground-based observations are unable 
to resolve a shell of that size.
There is no published finding chart of the quiescent object, so we
show, in Figure~\ref{fig9}, our \Ha on-band and off-band images, which 
clearly identify the nova.

The expansion velocity for CP~Cru has measured shortly after outburst
by \citet{dvb96} to be $\sim 2000$~km/s.

There are no previous distances estimate for CP~Cru.  Based on the
expansion velocity of 2000~km/s, a distance of 3.2~kpc is derived.  As
with HY~Lup (also with only \HST data), this distance may be an
overestimate to the true value.

The galactic extinction model yields a value $A_V = 1.98 \pm 0.74$.
The resulting absolute magnitude is $M_V = -5.3$, which is also 
an indication that the true maximum was missed. The nova has not yet 
returned to minimum light. It cannot be  used for maximum
or minimum luminosity determinations, unless additional observations become
available.

\subsection{Objects with no shell detected}

We failed to detect a shell around 15 objects, which are listed in
Table~\ref{tbl-4}. Three of the objects (GQ~Mus, V827~Her, and
V838~Her) were also observed with WFPC2 (program 7386), and no shell
was detected.  A shell around GQ~Mus was also not detected in
ground-based data by \citet{go98}.

Four objects --- V868~Cen, BY~Cir, V4361~Sgr, and N~Sco~97 --- do not
have published finding charts, and we show, in Figure~\ref{fig13}, our \Ha
on-band and off-band images, which clearly identify the quiescent
object.  \citet{rhh99} have identified a new candidate for V2104~Oph
based on positional coincidence (the field in the atlas of
\citet{due87b} is blank), and our image (Figure~\ref{fig13}) shows a faint
($V=20.50$) object, slightly \Ha bright, at the nominal position
(which is identical to the \citet{rhh99} candidate); the coordinates
of our candidate are 
$\rm \alpha_{2000} =  18^h 03^m25\fs 01$ and $\delta_{2000} =
+11^{\circ}47'51\farcs 1$ as measured from our image.

V888~Cen does not have a published chart, while \citet{due87b}
identifies candidates for V3645~Sgr and N~Car~1972. Our observations
of V888~Cen reveal that the marked object (the southern object of a
close pair) is \Ha bright, while for the latter two objects, there are
no \Ha-bright candidates (see Figure~\ref{fig13}).  Our observations
of N~Car~1971 (not shown) show a blank field at the nominal position
of the object.  However, examination of archival material
\citep{mac99} shows that the object was not detected on a direct plate
obtained 2~days after the discovery spectrum.  Thus, this object is
likely an emulsion defect, and also calls into question the reality of
N~Car~1972, which was found in a similar manner to N~Car~1971.

\section{Speed class and shape}

\citet{sod95} noted a correlation (at the $99\%$ confidence level)
between speed class ($t_3$) and the axial ratio of nova shells.  Such
a correlation is not unexpected, as the effects of common-envelope
evolution thought to occur during outbursts can remove varying amounts
of angular momentum from the system depending on the length of time
spent in this phase.  We have added our dataset to that of \citet{sod95} 
(for the three objects in common, we have averaged our ratio with theirs), 
and show, in Figure~\ref{fig14}, the merged plot.  We have fit a 
second-order polynomial to the data, and find that the data is
correlated at the $98\%$ level.  However, the fit is driven by 3
objects (DQ~Her, T~Aur, and RR~Pic), so the reality of the correlation
is still quite speculative.

\section{Absolute magnitudes of novae}

\subsection{Generalities}

In the previous section, distances to several well-investigated and
new novae, and their absolute magnitudes, were derived. We can use
these absolute magnitudes, in conjunction with the $t_2$ and $t_3$
times, to investigate a maximum magnitude-rate of decline (MMRD)
relationship.  There is no generally accepted light curve decline 
parameter; some authors
prefer $t_2$, to extend the MMRD to faint, extragalactic novae, but
the fluctuations of the light curve at maximum can easily reach
$2^{\rm m}$. The value of $t_3$ is much more stable in this
respect. Nevertheless, we will derive MMRD relations for both
parameters.

Another difference is the use of the $V$ (or visual) or the
$B$ (or photographic) light curve to derive the value of $t_i$. Since
novae are reddest at maximum, $t_{i,B}$ should be larger than
$t_{i,V}$. This is generally the case, but the general uncertainty of
pinning down the time and magnitude of maximum is so large that errors
in using the ``wrong'' color band are almost negligible. Since visual or
$V$ magnitudes are nowadays more common than blue magnitudes (which
were in the past often derived from photographic sky surveys), we will
restrict our analysis on $V$ magnitudes (see also Section~5.3). 

\subsection{Additional objects used for the study}

We have enhanced our sample of objects by adding expansion parallax
results from other authors.  For these objects, a critical evaluation on
assumed light curve data, expansion velocities and extinction values
was made.

\subsubsection{T Aur (1892)}

The most complete light curve of T Aur is based on Harvard
photographs, which show $m_{\rm pg} = 4.2$ at maximum, $t_2=45$ days,
$t_3=50$ days, and $m_{\rm pg,15} = 4.6$.  \citet{sod95} derived a distance
of $0.96\pm 0.22$~kpc, and \citet{gil94} derived an $A_V$ value of
$0.64$.  The absolute magnitude at maximum is $M_{\rm pg} = -6.6\pm
0.5$, $M_{\rm pg,15} = -6.2$, and the absolute magnitude at minimum is
$M_V=4.50$.

\subsubsection{GK Per (1901)}

The visual light curve was compiled by \citet{cam03}: $m_v = 0.19$ at
maximum, $t_2=7$ days, $t_3=13$ days, and $m_{v, 15} = 3.35$.
\citet{sod95} derived a distance of $0.455\pm 0.03$~kpc, and
\citet{wph89} derived an $E_{B-V}$ value of $0.3\pm 0.05$.  The
absolute magnitude is $M_V=-9.05\pm 0.3$ $M_{V,15}=-5.9$, and the 
absolute magnitude at minimum is $M_V=3.75$.

\subsubsection{V603 Aql (1918)}

The brightest nova of the 20th century reached $m_v=-1.1$ at maximum.
A detailed visual light curve \citep{cam19} gives single data, and
different averaged data points. These yield different decline rates,
and we adopt here $m_v = -1.1$ at maximum, $t_2 = 4$ days, $t_3 = 9$
days, and $m_{v,15} = 2.9$.  A value of $E_{B-V}$ of $0.07\pm 0.03$ was
derived by \citet{ver87}.  The expansion rate was determined by
\citet{due87a} to be $1\farcs 09$/yr, with an indication that the
rate decreases with time. With an expansion velocity
between $1700\pm 100$~km/s, a distance of $0.33$~kpc is
derived.  The absolute magnitude is $M_v=-8.9\pm 0.2$, $M_{v,15}= -4.9$, 
and the absolute magnitude at minimum is $M_V=3.90$.

\subsubsection{V476 Cyg (1920)}

The nova reached at maximum $m_v= 2.0$. A light curve by \citet{zak53}
yields $t_2 = 6$ days, $t_3 = 15$ days, and $m_{v,15} = 5.0$.  
\citet{sod95} derived a distance of $1.62\pm 0.12$ kpc, and a value of 
$A_V=0.82\pm 0.36$ was calculated from the galactic extinction model of
\citet{hak97}. The absolute magnitude at maximum is $M_{v}=-9.9\pm 0.6$, 
$M_{v,15}= -6.9$, and the absolute magnitude at minimum is
$M_V=5.85$.

\subsubsection{RR Pic (1925)}

The visual light curve, compiled by \citet{spe31}, reached $m_v =
1.18$, and has $t_2= 20$ days, $t_3 = 127$ days, and $m_{v,15} = 3.0$.
\citet{go98} derived a distance of $0.60\pm 0.06$~kpc from the ring
geometry and projection effects of the velocities. They assume an
inclination angle of $70^{\circ}$, and from the projected velocity
of $\pm 400$ km, they arrive at an intrinsic velocity of 850 km/s.

We derived a projected velocity of 405 km/s, an expansion rate of 
$0\farcs 157$/yr, and at a distance of 0.58~kpc, in good agreement 
with \citet{go98}. \citet{ver87} derived an $E_{B-V}$ value of 
$0.05\pm 0.03$. The absolute magnitude at maximum is $M_v=-7.8\pm 0.30$, 
$M_{v,15} = -6.0$, and the absolute magnitude at minimum is
$M_V=3.55$.

\subsubsection{XX Tau (1927)}

The photographic light curve, based on Harvard photographs, reached
$m_{\rm pg} = 6.0$, and has $t_2= 24$ days, $t_3 = 43$ days, 
and $m_{\rm pg, 15} = 7.4$.  \citet{coh85} derived an expansion rate of 
$\rm 0\farcs 039$/yr, and estimated an expansion velocity of 650~km/s, 
which yield a
distance of 3.5~kpc.  A value of $A_V=1.26\pm 0.57$ was derived from
the galactic extinction model of \citet{hak97}.  The absolute
magnitude at maximum is $M_{\rm pg}=-8.4\pm 0.75$, and $M_{\rm pg,15}
= -7.0$, and the absolute magnitude at minimum is
$M_V=4.60$.

\subsubsection{DQ Her (1934)}

A visual light curve, compiled by \citet{kuk37}, reached $m_v = 1.49$,
and has $t_2= 39$ days, $t_3 = 86$ days, and $m_{v,15} = 2.5$.
\citet{sod95} derived a distance of $0.40\pm 0.06$~kpc. A careful
study of \citet{hs92}, based on nebular slit spectra taken in the
1970s, yields a distance of $0.56\pm0.02$ kpc, while other recent
determinations, as quoted by \citet{hs92}, yield values near $0.48\pm
0.05$~kpc. We adopt the result of \citet{hs92}. \citet{ver87}
derived an $E_{B-V}$ value of $0.1\pm 0.05$, which is based on the
strength of the 2200\AA{} band and is a good average of other
determinations.  The absolute magnitude at maximum is $M_V=-7.5\pm
0.3$, $M_{V,15} = -6.55$, and the absolute magnitude at minimum is
$M_V=5.30$.

\subsubsection{CP Lac (1936)}

A visual light curve, compiled by \citet{ber45}, reached $m_v = 2.0$,
and has $t_2= 5.3$ days, $t_3 = 9.8$ days, and $m_{v,15} = 5.6$.
\citet{cr83} derived a distance of 1.35~kpc, while \citet{gil94}
derived an $A_V$ value of $0.64$. The galactic extinction model of
\citet{hak97} yields a value which is three times higher. 
The absolute magnitude at maximum is $M_V=-9.3$, $M_{V,15}=-5.7$,
and the absolute magnitude at minimum is $M_V=4.50$. All these 
values might be brighter by more than $1^{\rm m}$ if the extinction 
value based on UV spectroscopy is in error.

\subsubsection{BT Mon (1939)}

Maximum light was missed for this object.  \citet{go98} derived a
distance of $1.8\pm 0.3$~kpc, in perfect agreement with an earlier
result by \citet{mar83}. \citet{gil94} derived a value of $A_V=0.48$.
The observed maximum magnitude is $M_{pg} = -3.4\pm 0.35$, confirming
the fact that the nova was discovered long after maximum light.
The absolute magnitude at minimum is $M_V=3.85$.

\subsubsection{V450 Cyg (1942)}

A photographic light curve is based on observations of Ahnert, which
are scattered over various issues of the Beobachtungs-Zirkular of the
Astronomische Nachrichten. It reached $m_{\rm pg} = 7.5$, and has $t_2=
88$ days, $t_3 = 108$ days, and $m_{\rm pg, 15} \sim 8.1$.  \citet{sod95}
derived a distance of $3.5\pm 0.8$~kpc, and a value of $A_V = 1.3\pm
0.4$ is based on the galactic extinction model of \citet{hak97}. The
absolute magnitude at maximum is $M_{\rm pg}=-6.5\pm 0.6$, 
$M_{\rm pg,15} = -5.9\pm 0.6$. The magnitude at minimum is uncertain 
($\sim 19^{\rm m}$), and the absolute magnitude is about $+5$.

\subsubsection{V500 Aql (1943)}

A photographic light curve, based on Sonneberg and Harvard plates, peaks at
$m_{\rm pg, max} = 6.4$, and has $t_2 = 17$, $t_3 = 42$ days, and 
$m_{\rm pg, 15} = 8.3$. The nebular expansion parallax of \citet{coh85} 
yields a distance of 5.9 kpc, and a value $A_V = 0.94\pm 0.30$ is based 
on the galactic extinction model of \citet{hak97}. The absolute magnitude 
at maximum is $M_{\rm pg}=-8.7\pm 0.4$, $M_{\rm pg,15} = -6.8$.
\citet{szk94} derived $V=19.28$ as minimum magnitude, the 
absolute magnitude at minimum is $M_V=4.50$.

This nova fits nicely into the MMRD relations (see below). \citet{coh85} 
had assumed an $A_V$ of $3.0\pm 1.5$, which is unlikely because of the 
fairly blue color of the postnova. Her result, $M_V=-10.35$, was taken 
by \citet{del91} as evidence for including it in a hypothetical class of 
super-bright novae.

\subsubsection{DK Lac (1950)}

A photographic light curve is based on observations by \citet{lar54}.
Maximum is poorly covered: an observation of $m_{\rm pg}= 6.6$ belongs
to the rising branch of the light curve. Two days later, a poor observation 
gave $m_{\rm pg} = 5.35$:, and in the following days, the nova declined
smoothly by $0\fm 2$/day. Fitting the light curve of a fast nova 
through the points leads to a maximum magnitude $m_{\rm pg} = 5.2$. 
The light curve has $t_2=11$ days, $t_3=24$ days, and $m_{\rm pg, 15}=7.4$.  
\citet{sod95} derived a distance of $3.9\pm 0.5$~kpc, in good agreement 
with \citet{coh85}, and a value of $A_V=1.4\pm 0.4$ was derived from the
galactic extinction model of \citet{hak97}.  The absolute magnitude at
maximum is $M_{\rm pg} = -9.6\pm 0.5$, $M_{\rm pg,15} 
= - 7.4\pm 0.5$, and the absolute magnitude at minimum is $M_V=2.40\pm 0.6$. 
The nova reached a maximum of about $M_v = -9.85\pm 0.5$.

\subsubsection{V446 Her (1960)}

A photographic light curve, compiled by \citet{ber62}, reached $m_{\rm pg}
= 3.0$ on 1960 March 4. Outburst maximum is poorly covered by observations;
according to \citet{ber62}, spectroscopic evidence indicates that 
the real maximum of $2.50\pm 0.17$ was reached on March 1.
This assumption yields $t_2=7$ days, $t_3 = 11.5$ days, and $m_{\rm pg,15} =
5.85$.

\citet{coh85} derived a distance of 1.35~kpc, and a value of $A_V=1.12$ 
was derived by \citet{gil94}.  The absolute magnitude at maximum  is 
$M_{\rm pg}= -9.6\pm 0.2$, $M_{\rm pg, 15}= -6.5$, and the absolute magnitude 
at minimum is $M_V=5.85$ (quiescent light is superimposed by dwarf-nova
like outbursts). 

\subsubsection{V533 Her (1963)}

A photographic observation near maximum has $m_{pv} = 3.0$.  Follow-on
observations were compiled by \citet{alm66}, and we derive $t_2 = 22$
days, $t_3 = 46$ days, and $m_{V,15} = 4.4$.  \citet{sod95} derived a
distance of $0.56\pm 0.07$~kpc, based on a $v_{\rm exp} = 580$~km/s
as given by \citet{cr83}.
\citet{coh85}, using $v_{\rm exp} = 1050$~km/s, as determined from outburst
spectra, derived a distance of 1.32~kpc. 
Since nebular expansion velocities
may suffer inclination effects (see our discussion for CP Pup), 
the latter value appears more likely. Indeed recent observations
by \citet{go00} found a present-day $v_{\rm exp} = 850\pm 150$~km/s
and a new distance estimate of $1.25\pm 0.30$ kpc. \citet{ver87} derived an
$E_{B-V}$ value of $0.0\pm 0.04$, and \citet{gil94} a value
$A_V=0$. The new results lead to an absolute magnitude at maximum
$M_V= -7.5_{-0.45}^{+0.60}$,  $M_{V,15}=-6.1$, and the absolute 
magnitude at minimum is $M_V=4.60$.

\subsubsection{LV Vul (1968)}

A photoelectric light curve was compiled by \citet{due81}. The nova
reached $m_{V} = 4.6$, and has $t_2=21$ days, $t_3 = 43$ days, and
$m_{V,15} = 5.75$. \citet{sod95} derived a distance of $0.92\pm 0.08$~kpc, 
and a value of $A_V=1.75 \pm 0.42$ was derived from the galactic extinction
model of \citet{hak97}. The absolute magnitude at maximum is $M_V=-7.0\pm 
0.45$, $M_{V,15}=-5.85$, and the absolute magnitude at minimum is
$M_V=4.35$.

\subsubsection{V1229 Aql (1970)}

A light curve was published by \citet{cia74}. Maximum was
approximately $m_V=6.5$, $t_2=20$ days, $t_3 = 38$: days, and $m_{V,15} =
8.3$ for the photographic V light curve.  \citet{dd93} analyzed the
stellar profile and the structure of the \Ha + \NII-blend, and derived
an annual expansion rate of $ 0\farcs 072\pm 0\farcs 007$/yr, an expansion
velocity of 750~km/s, an average $E_{B-V}$ value of $0.50\pm 0.08$,
and a distance of $2.1\pm 0.9$~kpc. The absolute magnitude at maximum is
$M_V=-6.7^{+1.1}_{-0.8}$, $M_{V,15} = -4.9$. Since the postnova is a close
optical triple star, no accurate absolute magnitude at minimum can be given.

\subsubsection{FH Ser (1970)}

A photoelectric light curve was compiled by \citet{due96}. The nova
reached $m_{V} = 4.6$, and has $t_2=42$ days, $t_3 = 59$ days, and
$m_{V,15} = 5.45$.  \citet{sod95} derived a distance of $0.92\pm 0.13$~kpc, 
in good agreement with determinations of $0.87\pm0.09$~kpc
(\citet{dgb97}) and $0.85\pm 0.05$~kpc (\citet{due92}). 
New HST and ground-based observations \citep{go00} yield a distance of 
$0.95\pm 0.05$~kpc. Adopting this result and the $E_{B-V}=0.64\pm
0.16$ of \citet{dgb97}, we arrive at an absolute magnitude at maximum
of $M_V=-7.3\pm 0.5$, $M_{V,15} = -6.45$, and the absolute magnitude 
at minimum is $M_V=5.05$.

\subsubsection{V1500 Cyg (1975)}

A photoelectric light curve was compiled by \citet{due96}. The nova
reached $m_{V} = 1.78$, and has $t_2=2.35$ days, $t_3 = 3.7$ days, and
$m_{V,15} = 5.22$. \citet{sod95} derived a distance of $1.5\pm 0.2$~kpc, 
which is in good agreement with a previous estimate by \citet{wcj91}. The
$E_{B-V} = 0.50\pm 0.05$ was taken from \citet{lmu88}. The absolute
magnitude at maximum is $M_V=-10.7\pm 0.5$,  $M_{V,15} = -5.5$,
and the absolute magnitude at postoutburst minimum is $M_V=4.95$.

\subsection{The Maximum Magnitude-Rate of Decline Relationship}

The results of the preceding sections are summarized in Table~\ref{tbl-5}.
Some novae are only well-observed photographically, and the $m_{\rm pg, max}$
and $m_{pg,15}$ need to be transformed into the $V$ system. 
\citet{vdB87} find that novae at maximum have an intrinsic $\langle B-V
\rangle_0 = +0.23\pm 0.06$. A sample of nine objects with trustworthy 
reddening indicates a similar value, $\langle B-V \rangle_0 = +0.25\pm 0.05$.
The very fast nova V1500 Cyg is much bluer at maximum ($+0.05$), but the
objects considered here are all fairly slow, and were corrected
using the value $0\fm 25$. Since novae become bluer in later stages, we have
assumed $M_{V,15} = M_{\rm pg,15}$.

Table~\ref{tbl-5} also give the spectroscopic types of novae according to the
classification of \citet{wil92} (He/N, Fe II, and ``hybrid'' objects marked
as Fe/He/N), the light curve types (A, Ao, B, C, D), as defined 
by \citet{due81}, 
and mentions, for some recent objects, whether an object is a ONeMg nova.

The MMRD relations are shown in Figures~\ref{fig15}, \ref{fig16}, and 
\ref{fig17}. Most of the
fast, bright objects with smooth ``A-type'' light curves 
are He/N novae or hybrid objects, i.e. Fe II
novae that evolve into He/N novae. The three fastest pure Fe II novae are 
V476 Cyg, DK Lac and V351 Pup; they also show type A light curves, but 
their $t_3$ times are between 15 and 30 days, larger than those 
of He/N or hybrid novae ($t_3<15$ days). 
There appears to be no luminosity difference between 
the He/N novae and the Fe II novae, the only possible feature seems to be the
relative faintness of the `hybrid' objects relative to the `pure' ones.
All novae with structured light curves (types B,C,D) have $t_3 > 30$ days,
and they are, likely without exception, Fe II novae.

The ONeMg novae are thought to originate on massive white dwarfs, 
although their
luminosities appear to be consistent than the other novae of a
comparable rate of decline. The three objects in our diagram show 
light curve types A or B, and their $t_3$ times lie between 25 and 50 days.

If one sticks to the ``classical'' MMRD relations which hold for the ensemble
of points, the following statements can be made:

Using the $t_3$ time as the lightcurve parameter, the
relation appears almost linear,
$$
M_V = (-11.99\pm 0.56) + (2.54\pm 0.35) \log t_3.
$$
A similar relation results when using the $t_2$ time as the 
lightcurve parameter:
$$
M_V = (-11.32\pm 0.44) + (2.55\pm 0.32) \log t_2.
$$
In both cases, the residual for a single object is about $0\fm 6$, and a 
linear equation appears to be a good fit to the data.

The most recent fit of this type is by \citet{coh85}:
$$
M_V = (-10.70\pm 0.30) + (2.41\pm 0.23) \log t_2
$$
who, after selecting a ``high quality sample'' and correcting the other 
novae accordingly, was able to reduce the dispersion to $0\fm 47$.
It should be noted that a restriction on a near, less reddened sample,
or a correction of absolute magnitudes at maximum by assuming that
$M_{V,15} = const$ for all objects, as done by Cohen, does not lead
to a decisive improvement of the scatter in our data set, and thus
we refrain from any correction which may introduce systematic effects.
At an average $t_2 = 20$ days, Cohen's calibration is $0\fm 44$ fainter
than ours.

Another classical relation uses a specific stretched S-shaped curve, which is
apparent in samples of extragalactic novae. It was first suggested by 
\citet{vbp86}, and was recently revised by \citet{dvl95}. We have 
adopted their turn-over point, but have re-determined the zero-point and the 
amplitude:
$$
M_V = -8.02 -1.23 \arctan\frac{1.32-\log t_2}{0.23}.
$$
Our zero-point is almost identical to that of \citet{dvl95}, but the
``contrast'' between bright and faint novae has increased from $0\fm 81$ 
to $1\fm 23$. A possible explanation is that in the extragalactic sample
of  \citet{dvl95}, very bright and fast novae of the V1500 Cyg and CP Pup-type 
appear fainter because of insufficient sampling time, and the faintest
among the slow novae are missed (Malmquist bias).

Finally, it was shown by \citet{bdv55} that the absolute (pg) magnitude 
15 days after outburst is similar for novae of all speed classes. 
\citet{coh85} derived for 11 objects $M_{V,15} = -5.60\pm 0.45~\rm 
(s.e)$, while we obtain for 28 objects:
$$
M_{V,15} = -6.05\pm 0.44~\rm (s.e),
$$
indicating a similar zero-point difference ($0\fm 45$) between Cohen's and
our sample.
A regression shows that there is absolutely no trend in
$M_{V,15}$ versus, e.g., $t_3$-time, however, the
standard deviation ($0\fm 7$) is quite large.

There is one more way to look at the MMRD relation, if one takes
the light curve type as a second parameter. The dichotomy between the A/Ao
``super-Eddington'' and BCD ``Eddington'' groups, 
introduced in the investigation by \citet{due81}, appears well expressed. 
The ($M_V,t_2$) diagram can be divided in two fields: the A/Ao group
has $M_V \le 8^{\rm m}$ and $\log t_2 < 1.20$, the BCD group
has $M_V \ge 8^{\rm m}$ and $\log t_2 \ge 1.20$ (two poorly investigated
objects, V500 Aql and XX Tau, may have poorly determined luminosities and/or
decline times, and are not considered in the following discussion). 
In the ($M_V,t_3$) diagram, the two groups are separated as well; 
the dividing line in time is now at $\log t_3 = 1.5$. 

If one examines the ``super-Eddington'' and ``Eddington'' groups separately, 
MMRD relations are quite weakly expressed; in the $M_V,t_3$ diagram, the
A/Ao group follows the relation
$$
M_V = (-11.26\pm 0.84)  + (1.58\pm 0.78) \log t_3
$$
with a correlation coefficient of 0.61. The BCD group shows only a weak 
correlation coefficient of 0.21 between magnitude and decline time;
$\langle M_V\rangle = -7.09 \pm 0.61 (s.e.)$. A linear relation for 
these objects is:
$$
M_V = (-8.13\pm 1.26) + (0.57\pm 0.68) \log t_3.
$$

In the $M_V,t_2$ diagram, the A/Ao group follows the relation
$$
M_V = (-10.79\pm 0.92) + (1.53\pm 1.15) \log t_2 
$$
with a correlation coefficient of 0.45, and the BCD group yields
$$
M_V = (-8.71\pm 0.82) + (1.03\pm 0.51) \log t_2
$$
with a correlation coefficient of 0.46 (of course, the $\langle M_V\rangle =
-7.09$ is the same). Such a type of fit in two separate
regions mimics the stretched S-shaped curve used for a fit of the complete 
sample, but in the present case we have a criterion (the light curve form) 
to separate two regions, and to use different linear relations (or simple
averages) in them.

The overall scatter in the MMRD relation -- which is of the order of
$0\fm 5$ in all these relations -- may indicate the presence of hidden
second order parameters (spectroscopic type, light curve type, mass of
underlying white dwarf\dots ) which should influence the luminosity. 
However, the individual errors are too large, so the sample is not able to
indicate which parameter is most decisive. Thus,
the MMRD relation can be used as a reliable distance 
indicator only in a statistical way; for a
single object, the absolute magnitude can be off by $\sim 0\fm 5$.
If samples are taken from different stellar populations, systematic
luminosity differences may be present. This does not devaluate the use
of novae as distance indicators, it simply means that caution must be
used when such relations are applied (similar problems occur in other
distance indicators like supernovae, Cepheids, and RR Lyrae stars). A
better understanding of the MMRD relation is also the basis of a
better understanding of the physics of nova eruptions.

\subsection{Novae at minimum}

Another result of our study is the derivation of absolute magnitudes of
classical novae after eruption. Table \ref{tbl-6} lists the absolute
visual magnitudes of novae, which have settled to a more or less constant
minimum brightness. Some of the objects have known orbital periods and
inclination angles; relevant data were taken from the catalog of
\citet{rit98}, and these magnitudes $M_V$ were corrected for inclination
effects, using the formula given by \citet{war87}; the corrected magnitudes
$M_V^{\rm corr}$ are also listed. All novae for which no inclination
information was available, were corrected by assuming an average inclination
$i=46^\circ$. Recent results on orbital periods are available for V500 Aql
\citep{hae99} and V446~Her and V533~Her \citep{tho00}.

In addition to the nova data, absolute magnitudes of dwarf novae
at maximum, based on HST FGS parallaxes, and of novalike stars, mainly based
on Hipparcos parallaxes, were derived (the absolute magnitudes supersede those
given by \citet{due99}). Results are summarized in Fig.~\ref{fig18}.
If one discards novae in or below the period gap (CP Pup, RW UMi, QU Vul) and
the long-period system GK Per, as well as the peculiar novalike AE Aqr,
the remaining systems (with $-0.9 <\log P~[{\rm days}] < -0.4$) indicate a
dependence of absolute magnitude on period:
$$
M_V = (0.59\pm 1.74) - (5.15\pm 2.41)\log P\, \rm [days] 
$$ 
(correlation coefficient 0.47). 
If the period (in hours) is used, the relation is
$$
M_V = (6.60\pm 1.0) - (0.49\pm 0.20) P\, \rm [hours]
$$
which, at $P=5$ hours, yields absolute magnitudes which are $0\fm 3$ brighter
than those predicted \citet{war95}'s formula for dwarf novae at maximum.

Summing up, no obvious separation of absolute magnitudes of objects belonging to
different groups (novae at minimum, dwarf novae at outburst, and novalike
systems at normal light) can be seen. This again shows the intimate
connection of these groups, which is also obvious from space density
considerations, and the hibernation scenario, which indicates that only
by including novalike stars and dwarf novae to the space densities of novae,
realistic recurrence times of novae can be obtained \citep{due95}.

\section{Summary}

We have obtained optical imaging of 30 recent novae, and have detected
shells, in combination with \HST data, for 14 objects.  We fail to
detect the previously observed shell around RW~UMi, while PW~Vul,
V1974~Cyg, and V842~Cen
appear to show two distinct shells.  Expansion parallaxes for the
novae have been derived, and combined with revised
distances for other novae, have been used to derive absolute
magnitudes.  A study of various maximum magnitude-rate of decline relations
is in general agreement with previous ``linear'' determinations. 
However, a relation making use of the light curve type may yield a somewhat
better fit to the data. It clearly shows the existence of two types
of novae: (1) The fast ($t_2<13, t_3<30$ days) 
super-Eddington novae, which may be He/N, ``hybrid''
Fe II novae, and, if their evolution is somewhat slower, Fe II novae.
They all show smooth light curves with well-defined maxima (A), and some
of them may show quasi-periodic light oscillations in their later 
evolution (Ao). (2) the slow $t_2>13,~t_3>30$ days) Eddington-novae with 
structured light 
curves (Ba), double maxima (Bb), standstills at maximum (Ca), dust formations
at later stages (Ca, Cb), and long premaximum stages (D), 
seem to belong exclusively to the Fe II spectroscopic type.
The merging of both groups leads to a well-expressed MMRD; if both
groups are considered separately, their MMRDs show smaller slopes,
and a jump of $\sim 2^{\rm m}$ occurs at $t_2 = 13$ days and $t_3 = 30$ days.
At minimum, most novae have similar magnitudes as those of dwarf novae at
maximum light and novalike stars.

\acknowledgments

We wish to thank the AAVSO for providing data from their International
Database, which is based on observations submitted to the AAVSO by
variable star observers worldwide, and the AFOEV for providing data
from their internet database. We also thank Tim Naylor and Fred
Ringwald for providing a copy of their PW~Vul data, Jeff Robertson for
providing a copy of his V2014~Oph data, Jack MacConnell for
investigating the reality of Nova~Car~1971 and 1972, and Pierluigi
Selvelli for information on interstellar extinction data. HWD also
gratefully acknowledges the hospitality and support of Space Telescope
Science Institute, where he could carry out part of the work on which
this investigation is based.  We wish to thank an anonymous referee
for many useful comments on this manuscript.

\clearpage

\figcaption[fig1.ps]{\Ha and \OIII images of CP Pup 1942 -- flat-fielded
images and contour plots. \label{fig1}}

\figcaption[fig2.ps]{\Ha images of CT Ser 1948 -- flat-fielded
image, radial profile, and PSF-subtracted data. \label{fig2}}

\figcaption[fig3.ps]{\Ha contour plots of RW UMi 1956 -- both the nova 
and a nearby field star of comparable brightness.  There is no indication of
a shell. \label{fig3}}

\figcaption[fig4a.ps,fig4b.ps]{(a) \Ha and (b) \OIII images of HR Del 1967 --
flat-fielded images, radial profiles, and PSF-subtracted
data. \label{fig4}}

\figcaption[fig5.ps]{\Ha images of V3888 Sgr 1974 -- flat-fielded
image and contour plots for the direct and PSF-subtracted data. \label{fig5}}

\figcaption[fig6.ps]{\Ha images of NQ Vul 1976 -- flat-fielded
image and contour plots for the direct and PSF-subtracted data. \label{fig6}}

\figcaption[fig7.ps]{\Ha images of PW Vul 1984 -- flat-fielded
image, radial profile, and PSF-subtracted data. \label{fig7}}

\figcaption[fig8a.ps,fig8b.ps]{(a) \Ha and (b) \OIII images of QU Vul 1984 --
flat-fielded images, radial profiles, and PSF-subtracted data. While
the [O III] PSF-subtracted image shows the object weakly, the radial
profile clearly demonstrates the extended nature of the nova. \label{fig8}}

\figcaption[fig9.ps]{\Ha on-band and off-band images of novae with
shells detected by \HzST. The objects are clearly \Ha bright. \label{fig9}}

\figcaption[fig10a.ps,fig10b.ps]{(a) \Ha and (b) \OIII images of V842 Cen
1986 -- flat-fielded images, radial profiles, and PSF-subtracted
data. \label{fig10}}

\figcaption[fig11.ps]{\Ha images (on-band and off-band) of V351
Pup 1991 -- flat-fielded images, radial profiles, and PSF-subtracted
data. While the PSF-subtracted image fails to detect any shell, the
radial profile of the nova is clearly extended in the on-band image,
while it is not in the off-band image. \label{fig11}}

\figcaption[fig12a.ps,fig12b.ps]{(a) \Ha and (b) \OIII images of V1974 Cyg
1992 -- flat-fielded images, radial profiles, and PSF-subtracted
data. \label{fig12}}

\figcaption[fig13a.ps,fig13b.ps]{\Ha on-band and off-band images of
novae without published charts. For V868 Cen, BY Cir, V4361 Sgr, and N
Sco 1997, the object is clearly \Ha bright. For V888 Cen, the southern
object is \Ha bright, although blending with the northern companion masks
this brightness. \label{fig13}}

\figcaption[fig14.ps]{Plot of the axial ratio of the nova shells
versus t$_{3}$.  The points have not been corrected for inclination.
Points 1-9 are from this work, while the remainder are taken from
\citet{sod95}.  The values for CP Pup, HR Del, and NQ Vul are averages
of our data and that of \citet{sod95}.  Note that the time for CT Ser is
t$_{2}$ rather than t$_{3}$. \label{fig14}}

\figcaption[fig15.ps]{Linear MMRD relation for $t_3$. The dashed line is
the global fit to the data, while the solid lines are the fits for the
``super-Eddington'' (upper left) and ``Eddington'' (lower right) novae.  The
objects XX~Tau and V500~Aql, situated in the upper right area of the diagram, 
are poorly studied, and may have uncertain
luminosities and decline times. The average error of a data point is indicated
in the upper right corner. \label{fig15}}

\figcaption[fig16.ps]{Same as figure~\ref{fig15} for $t_2$. \label{fig16}}

\figcaption[fig17.ps]{Curved MMRD relation for the decline rate $v_d$,
i.e. the average daily decline in magnitudes from maximum to maximum + 
$2^{\rm m}$. The average error of a data point is indicated
in the upper right corner. \label{fig17}}

\figcaption[fig18.ps]{Absolute visual magnitudes for classical novae at 
minimum (filled circles), for dwarf novae at maximum (open circles), 
and for novalike stars at normal light (open squares).
Systems in or below the period gap, the long-period nova system GK Per,
and the peculiar novalike object AE Aqr were not included in the
fit.\label{fig18}}

\clearpage

\renewcommand{\arraystretch}{0.6}
\pagestyle{empty}

\begin{deluxetable}{rccrrc}
\footnotesize
\tablecaption{Summary of Observations \label{tbl-1}}
\tablewidth{0pt}
\tablehead{
\colhead{ } & \colhead{ }   & \colhead{ }   &
\colhead{Exp. Time} &  \colhead{Exp. Time}   & \colhead{ } \\
\colhead{Object} & \colhead{Telescope}   & \colhead{Date}   &
\colhead{~~~~~~~~~\Hza} & \colhead{~~~~~~~[O III]} & \colhead{V} 

}

\startdata

GQ Mus    & ESO Dutch-0.9m & 21 March 1998 &  900s & 1200s & 18.37 \\
V351 Pup  &                &               &  600s & 1200s & 19.55 \\
HY Lup    &                &               &  600s & 1200s & 18.86 \\
CP Cru    &                &               &  600s & 1200s & 19.48 \\
N Car 72  &                &               & 1200s &       & unident. \\
V365 Car  &                & 22 March 1998 &  600s &       & 18.47 \\
V868 Cen  &                &               &  600s &  900s & 19.88 \\
V888 Cen  &                &               &  600s &  900s & 16.38 \\
CP Pup    &                &               &  600s &       & 14.99 \\
V842 Cen  &                & 23 March 1998 &  450s &  600s & 15.82 \\
V812 Cen  &                &               &  600s &  900s & 20.44 \\
BY Cir    &                &               &  450s &  600s & 15.86 \\
V4361 Sgr &                &               &  180s &  180s & 16.85 \\
N Sco 97  &                &               &  180s &  180s & 17.58 \\
CP Pup    &                &               &       & 1200s &       \\
N Car 71  &                &               &  900s &       &  unident.  \\
\\
RW UMi    & KPNO-2.1m      & 28 May 1998   & 1050s & 1050s & 18.95 \\
V827 Her  &                &               &  720s &  600s & 18.08 \\
HR Del    &                &               &   60s &   60s & 12.16 \\
V1819 Cyg &                &               &       &  600s & 20.33 \\
\\
CT Ser    &                & 29 May 1998   & 1050s & 1050s &   - \\
V794 Oph  &                &               & 1050s & 1050s &   - \\
V838 Her  &                &               & 660s & 900s   & 19.07 \\
QV Vul    &                &               & 660s & 900s   & 18.61 \\
V1819 Cyg &                &               & 660s &        &  \\
V1974 Cyg &                & 31 May 1998   & 660s & 660s   & 16.06 \\
PW Vul    &                &               &      & 360s   & 17.51 \\
V2104 Oph &                &               & 3600s& 2700s  & 20.50 \\
V3888 Sgr &                & 01 June 1998  & 3600s&        & 20.96 \\
V3645 Sgr &                &               & 150s &        & unident. \\
NQ Vul    &                &               & 660s & 600s   & 17.70 \\
PW Vul    &                &               & 360s &        &  \\
\\
QU Vul    & KPNO-0.9m      & 30 June 1998  & 1550s & 1800s & 18.31 \\

\enddata

\end{deluxetable}

\clearpage

\begin{deluxetable}{lcrcccc}
\footnotesize
\tablecaption{Systems with Shells \label{tbl-2}}
\tablewidth{0pt}
\tablehead{
\colhead{Nova} & \colhead{Measurement\tablenotemark{a}} & \colhead{Diameter}   &
\colhead{Year} & \colhead{Ref.} & \colhead{v$_{exp}$} &
\colhead{Ref.} \\
\colhead{(Outburst date)} & \colhead{Technique} & \colhead{(\arcsec)}   & \colhead{Taken} &
\colhead{ } & \colhead{(km/s)}   & \colhead{ } 
}

\startdata

CP Pup (1942.86)    & D &              16.9 & 1998.23 &   0 & 815 & 12 \\
                    & D &              16.0 & 1995.75 &   1 &  \\
                    & D &              13.4 & 1987.20 &  24 &  \\
CT Ser (1948.27)    & D &   $7.9\times 7.9$ & 1998.23 &   0 &  535 & 8 \\
RW UMi (1956.73)    &   &          not seen & 1998.41 &   0 &  950 & 11 \\
                    &   &          not seen & 1997.98 &   1 & \\
                    & D &               3.0 & 1993.69 &  27 &  \\
                    & R &               2.0 & 1983.5  &  11 &  \\
HR Del (1967.52)    & D &   $9.8\times 8.5$ & 1998.41 &   0 &470,540 &32,34 \\
                    & D &   $8.7\times 6.9$ & 1997.40 &   1 &520,560 & 8,35 \\
                    & D &  $11.5\times 8.5$ & 1993.69 &  27 &  \\
                    & D &               5.5 & 1992.58 &   7 &  \\
                    & D &   $3.7\times 2.5$ & 1981.33 &  13 &  \\
V3888 Sgr (1974.76) & D & $5.2\times 4.6$ & 1998.41 & 0 & 1300 & 22 \\
NQ Vul (1976.81)    & D &   $7.3\times 7.0$ & 1998.41 &   0 & 1300,1700 &15,16 \\
                    &   &          not seen & 1997.92 &   1 & 950,705 &14,8 \\
                    & D &                 8 & 1993.69 &  27 & \\
PW Vul (1984.57)    & D &   $1.5\times 1.5$ & 1998.79 &   1 & 1200,470 &17,3 \\
                    & D &   $4.0\times 3.7$ & 1998.41 &   0 &  \\
                    & R &               1.1 & 1993.64 &   3 & \\
QU Vul (1984.99)    & D &   $1.6\times 1.7$ & 1998.71 & 1,6 & 790,1200 &19,26 \\
                    & D &   $5.6\times 5.6$ & 1998.50 &   0 & 570,1190 &37,4 \\
                    & D &   $1.6\times 1.3$ & 1998.18\tablenotemark{e} & 2 \\
                    & R &               0.8 & 1996.35\tablenotemark{b} & 5 \\
                    & R &               1.9 & 1994.52 &   4 &  \\
                    & D &   $0.2\times 0.1$ & 1986.34\tablenotemark{d} &  26 &  \\
V1819 Cyg (1986.57) & D &   $0.5\times 0.5$ & 1999.18\tablenotemark{c} &   1 & 600 & 38 \\
                    &   &          not seen & 1998.41 &   0 &  700   & 39 \\
V842 Cen (1986.89)  & D &   $5.6\times 6.0$ & 1998.23 &   0 &  535   & 23 \\
                    & D &               1.6 & 1995.15 &  10 &  525   & 33 \\
QV Vul (1987.87)    & D &   $0.8\times 0.9$ & 1998.95\tablenotemark{c} & 1 & 515 & 37 \\
                    &   &           visible & 1998.95 &   1 & \\
                    & D &   $1.4\times 1.4$ & 1998.62\tablenotemark{e} & 2  \\
                    &   &          not seen & 1998.41 &   0 & \\
V351 Pup (1991.99)~~~~~~~~  & D &               1.3 & 1998.23 & 1,6 & 2000 & 36 \\
                    &   &           visible & 1998.12 &   0 & 1200 & 40 \\
                    & R &              0.58 & 1995.17 &  36 & \\
V1974 Cyg (1992.14) & D & $8.6\times 9.0$   & 1998.41 &   0 & 825,1100 &9,20\\
                    & D &   $1.8\times 1.7$ & 1998.12 & 1   & 1500,860 &21,28\\
                    & D &   $1.4\times 1.3$ & 1998.11\tablenotemark{e} & 2 \\
                    & D &   $0.4\times 0.5$ & 1994.38\tablenotemark{f} & 30 \\
                    & D &   $0.3\times 0.4$ & 1994.03\tablenotemark{f} & 30 \\
                    & D &   $0.2\times 0.3$ & 1993.41\tablenotemark{g} & 31 \\
HY Lup (1993.71)~~~~~~~~~~~~~~~~~ & D &   $0.5\times 0.5$ & 1999.12 &   1 &     2700,400 & 25,18 \\
                    &   &          not seen & 1998.23 &   0 \\
CP Cru (1996.65)    & D &   $0.6\times 0.6$ & 1998.91 & 1,6 &     2000 & 29 \\
                    &   &          not seen & 1998.23 &   0 \\
\enddata

\tablenotetext{a}{D = direct image, R = radial profile fit}
\tablenotetext{b}{$2.2\micron$ data}
\tablenotetext{c}{HST - F658N ([N II]$\lambda6583$) data}
\tablenotetext{d}{15 GHz data}
\tablenotetext{e}{HST - F222M ($2.15-2.30\micron$) data}
\tablenotetext{f}{HST - F501N ([O III]$\lambda5007$) data}
\tablenotetext{g}{HST - F278M (Mg II~$\lambda2800$) data}

\tablerefs{
(0) this work; (1) HST WFPC2 \Ha image; (2) HST NICMOS image; (3) \citet{rn96};
(4) \citet{dgb97}; (5) \citet{shi98}; (6) \citet{rwo98}; (7) \citet{sod94}; 
(8) \citet{cr83}; (9) \citet{ros96}; (10) \citet{go98}; (11) \citet{coh85}; 
(12) \citet{gra53}; (13) \citet{koh81}; (14) \citet{kw78};(15) \citet{cs78};
(16) Younger(1980); (17) \citet{kw86}; (18) \citet{due00}; (19) \citet{ri87}; 
(20) \citet{cgp97}; (21) \citet{sho93}; (22) \citet{lwv76}; (23) \citet{sek89};
(24) \citet{due87c}; (25) \citet{del93}; (26) \citet{tshp87}; 
(27) \citet{sod95}; (28) \citet{ckh95}; (29) \citet{dvb96}; 
(30) \citet{plhk95}; (31) \citet{par94}; (32) \citet{mal75}; 
(33) \citet{sei90}; (34)  \citet{koh81}; (35) \citet{solf83};
(36) \citet{obd96}; (37) \citet{and91}; (38) \citet{whi89}; (39) \citet{and89};
(40) \citet{ps95}.
}

\end{deluxetable}

\clearpage

\begin{deluxetable}{lcccccc}
\footnotesize
\tablecaption{Nova Distances \label{tbl-3}}
\tablewidth{0pt}
\tablehead{
\colhead{Nova} & \colhead{Age}   & \colhead{v$_{exp}$} & \colhead{Exp. Rate} &
\colhead{Distance} 
\\
\colhead{ } & \colhead{(yr)} &  \colhead{adopted} & \colhead{(\arcsec/yr)}   &
\colhead{(pc)} 
}

\startdata

CP Pup    & 55.37 &  815 &   0.153 & 1120 \\
          & 52.89 &      &   0.151 & 1140 \\
          & 44.34 &      &   0.151 & 1140 \\
CT Ser    & 49.96 &  535 &   0.079 & 1430 \\
RW UMi    & 36.96 &  950 &   0.037 & 4900 \\
          & 26.8  &      &   0.041 & 5400 \\
HR Del    & 30.89 &  525 &   0.149 &  750 \\
          & 29.88 &      &   0.146 &  760 \\
          & 26.17 &      &   0.220 &  505 \\
          & 25.06 &      &   0.110 & 1010 \\
          & 13.81 &      &   0.134 &  825 \\
V3888 Sgr & 23.65 & 1300 &   0.110 & 2495 \\
NQ Vul    & 21.60 & 1025 &   0.169 & 1280 \\
          & 16.88 & 1025 &   0.237 & 910 \\
          & 16.88 &  545 &   0.090 & 1280 \\
PW Vul    & 14.22 &  470 &   0.053 & 1880 \\
          & 13.84 & 1200 &   0.145 & 1750 \\
          &  9.07 &  470 &   0.061 & 1635 \\
QU Vul    & 13.72 &  570 &   0.060 & 2005 \\
          & 13.51 & 1500 &   0.207 & 1530 \\
V1819 Cyg & 12.61 &  700 &   0.020 & 7385 \\
V842 Cen  & 11.34 & (1600) &   0.255 & (1325) \\
          &  8.26 &  525 &   0.097 & 1140 \\
QV Vul    & 11.08 &  515 &   0.041 & 2675 \\
V351 Pup  &  6.24 & 1200 &   0.100 & 2530 \\
          &  3.18 &      &   0.091 & 2775 \\
V1974 Cyg &  6.27 & (4500) &   0.718 & (1320) \\
          &  2.24 & 1060 &   0.112 & 2000 \\
          &  1.89 &      &   0.106 & 2115 \\
          &  1.30 &      &   0.115 & 1940 \\
HY Lup    &  5.41 &  400 &   0.046 & 1800 \\
CP Cru    &  2.26 & 2000 &   0.133 & 3180 \\

\enddata

\end{deluxetable}

\clearpage

\begin{deluxetable}{rccl}
\footnotesize
\tablecaption{Systems with no resolved shell detected at \Ha \label{tbl-4}}
\tablewidth{0pt}
\tablehead{
\colhead{Nova} & \colhead{Outburst Year} & \colhead{Age}  & 
\colhead{Limiting (3$\sigma$) Flux\tablenotemark{a}} \\
\colhead{ } & \colhead{ } & \colhead{(years)}  & 
\colhead{Ground~~~~~~~WFPC2} 
}

\startdata

~~~~~~~V794 Oph & 1939.45 & 58.78 & ~~4.95 \\
V365 Car & 1948.74 & 49.49 & ~~52.2 \\
V812 Cen & 1973.25 & 24.98  & ~~69.0 \\
V2104 Oph& 1976.76 & 21.68  & ~~0.44 \\          
GQ Mus   & 1983.05 & 15.18  & ~~26.4~~~~~~~~~~~2.57 \\
V827 Her & 1987.05 & 12.41  & ~~0.86~~~~~~~~~~~1.32 \\
V838 Her & 1991.23 &  7.18  & ~~2.05~~~~~~~~~~~1.17 \\
V868 Cen & 1991.25 & 6.98  & ~~32.1 \\
BY Cir   & 1995.07 & 3.16  & ~~\tablenotemark{b} \\
V888 Cen & 1995.15 & 3.08  & ~~\tablenotemark{b} \\
V4361 Sgr& 1996.53 & 1.70  & ~~\tablenotemark{b} \\
N Sco 1997   & 1997.43 & 0.80  & ~~\tablenotemark{b} \\

\enddata

\tablenotetext{a}{$10^{-15}$ erg cm$^{-2}$ s$^{-1}$ arcsec$^{-2}$}
\tablenotetext{b}{too young to be resolved}

\end{deluxetable}

\clearpage

\begin{deluxetable}{lcccccccl}
\footnotesize
\tablecaption{Nova Luminosities \label{tbl-5}}
\tablewidth{0pt}
\tablehead{
\colhead{Object} & \colhead{spec.} & \colhead{light curve} & \colhead{$t_2$} 
& \colhead{$t_3$} & \colhead{decline rate} & \colhead{$M_{\rm max}$} &
\colhead{$M_{15}$} & \colhead{comment} 
\\
\colhead{ } & \colhead{type\tablenotemark{a}} & \colhead{type} 
& \colhead{(days)} & \colhead{(days)} 
& \colhead{$^m$/day} & \colhead{$V$} & \colhead{$V$} & \colhead{ }
}

\startdata
V500 Aql & He?& Ao & 17 & 42 & 0.118 & $-8.7\pm 0.4$ & $-6.8 $& pg \\
V603 Aql & hy & Ao &  4 &  9 & 0.500 & $-8.9\pm 0.2$ & $-4.9 $& \\
V1229 Aql& Fe & B? & 20 & 38 & 0.100 & $-6.7\pm 0.95$ & $-4.9 $& \\
T Aur    & Fe & Ca & 45 & 50 & 0.044 & $-6.8\pm 0.5$ & $-6.2 $& pg \\
V842 Cen & Fe & Cb & 35 & 48 & 0.057 & $-7.4\pm 1.0$ & $-6.5 $& \\
V450 Cyg & Fe & Ca & 88 &108 & 0.023 & $-6.5\pm 0.6$ & $-5.9 $& \\
V476 Cyg & Fe & A  &  6 & 15 & 0.333 & $-9.9\pm 0.6$ & $-6.9 $& \\
V1500 Cyg& hy & A  & 2.4& 3.7& 1.111 &$-10.7\pm 0.5$ & $-5.5 $& \\
V1819 Cyg& Fe & Bb & 37 & 89 & 0.054 & $-6.8\pm 0.5$ & $-5.6 $& \\ 
V1974 Cyg& Fe & Ba & 17 & 37 & 0.125 & $-8.0\pm 0.1$ & $-6.4 $& ONeMg\\ 
HR Del   & Fe & D  &172 &230 & 0.012 & $-6.1\pm 0.4$ & $-4.8 $& \\
DQ Her   & Fe & Ca & 39 & 86 & 0.051 & $-7.5\pm 0.3$ & $-6.55$& \\
V446 Her & He & A  &  7 &11.5& 0.426 & $-9.9\pm 0.2$ & $-6.5$& pg \\
V533 Her & Fe & Ba & 22 & 46 & 0.091 & $-7.5\pm 0.5$ & $-6.1 $& \\
CP Lac   & hy & A  & 5.3& 9.8& 0.377 & $-9.3       $ & $-5.7 $& \\
DK Lac   & Fe & Ao?& 11 & 24 & 0.182 & $-9.6\pm 0.5$ & $-7.4 $& \\
GK Per   & He & Ao &  7 & 13 & 0.286 & $-9.0\pm 0.3$ & $-5.9 $& \\
RR Pic   & Fe & D  & 20 &127 & 0.100 & $-7.8\pm 0.3$ & $-6.0 $& \\ 
CP Pup   & He & A  &  6 &  8 & 0.333 & $-10.7\pm 0.5$& $-6.1 $& \\
V351 Pup & Fe & A  & 10 & 26 & 0.200 & $-8.4\pm 0.5$ & $-6.1 $& ONeMg\\
FH Ser   & Fe & Cb & 42 & 59 & 0.048 & $-7.3\pm 0.5$ & $-6.45$& \\
XX Tau   & Fe & Cb & 24 & 43 & 0.083 & $-8.4\pm 0.75$& $-7.0 $& pg \\
RW UMi   & ?  & Ba & 48 & 88 & 0.042 & $-7.8\pm 0.9$ & $-7.3 $& pg \\
LV Vul   & Fe & Ba & 21 & 43 & 0.095 & $-7.0\pm 0.45$& $-5.85$& \\
NQ Vul   & Fe & Bb & 23 & 53 & 0.087 & $-6.1\pm 0.8$ & $-4.85$& \\
PW Vul   & Fe & Bb & 82 &126 & 0.024 & $-6.7\pm 0.4$ & $-5.3 $& \\
QU Vul   & Fe & Ba & 22 & 49 & 0.091 & $-7.5\pm 0.25$& $-6.35$& ONeMg\\
QV Vul   & Fe & Cb & 50 & 53 & 0.040 & $-6.4       $ & $-5.45$& \\
\enddata

\tablenotetext{a}{He = He/N type, Fe = Fe II type, hy = hybrid type, 
evolving from Fe II into He/N type}

\end{deluxetable}

\begin{deluxetable}{lcccccccl}
\footnotesize
\tablecaption{Cataclysmic Variable Luminosities\label{tbl-6}}
\tablewidth{0pt}
\tablehead{
\colhead{Nova} & \colhead{P(orbit)} & \colhead{$V_{\rm min}$} 
& \colhead{distance} 
& \colhead{$A_V$} & \colhead{$M_V$} & \colhead{$i$} 
& \colhead{$M_V^{\rm corr}$} & \colhead{comment}
\\
\colhead{ } & \colhead{[days]} & \colhead{} 
& \colhead{[kpc]} & \colhead{} 
& \colhead{} & \colhead{[degrees]} & \colhead{} & \colhead{} 
}

\startdata
V500 Aql & 0.1452 & 19.25 & 5.9  & $0.94\pm 0.30$ & 4.50 &    & 4.85 & Novae \\
V603 Aql & 0.1381 & 11.70 & 0.33 & $0.22\pm 0.10$ & 3.90 & 17.& 4.80 & 
at minimum\\
V1229 Aql&        &       & 2.1  & $1.58\pm 0.25$ &      &    &      & \\
T Aur    & 0.2044 & 15.03 & 0.96 & $0.64        $ & 4.50 & 57.& 4.45 & \\
V842 Cen &        & 15.82 & 1.15 & $1.73\pm 0.16$ &      &    &      & \\
V450 Cyg &        & 19.   & 3.5  & $1.3 \pm  0.4$ & 5.:  &    & 5.4  & \\
V476 Cyg &        & 17.70 & 1.62 & $0.82\pm 0.36$ & 5.85 &    & 6.20 & \\
V1500 Cyg& 0.1396 & 17.42 & 1.5  & $ 1.6\pm 0.16$ & 4.95 & ?  & 5.35 & \\
V1819 Cyg&        & 20.33 & 7.4  & $1.10\pm 0.47$ & 4.90 &    & 5.25 & \\
%1974 Cyg& 0.0813 & 16.06 & 1.8  & $1.10        $ &      &    &      & \\
HR Del   & 0.2142 & 12.16 & 0.76 & $0.47\pm 0.10$ & 2.30 & 40.& 2.80 & \\
DQ Her   & 0.1936 & 14.38 & 0.48 & $0.32\pm 0.16$ & 5.30 &86.5& 2.70 & \\
V446 Her & 0.2071 & 17.61 & 1.35 & $1.12        $ & 5.85 &    & 6.20 & \\
V533 Her & 0.1471 & 15.06 & 1.25 & $0.00\pm 0.04$ & 4.60 & ?  & 4.95 & \\
CP Lac   &        & 15.80 & 1.35 & $0.64        $ & 4.50 &    & 4.90 & \\
DK Lac   &        & 16.75 & 3.9  & $1.4 \pm  0.4$ & 2.40 &    & 2.75 & \\
BT Mon   & 0.3338 & 15.59 & 1.8  & $0.48        $ & 3.85 & 82 & 1.90 & \\ 
GK Per   & 1.9968 & 13.01 & 0.455& $0.95\pm 0.16$ & 3.75 &$<73$&4.15 & \\
RR Pic   & 0.1450 & 12.50 & 0.58 & $0.16\pm 0.10$ & 3.55 & 65 & 3.45 & \\
CP Pup   & 0.0614 & 14.99 & 1.7  & $0.79        $ & 3.90 & 40:& 4.45 & \\
V351 Pup &        & 19.55 &      & $2.29\pm 0.32$ &      &    &      & \\
V3888 Sgr&        & 20.96 & 2.5  & $3.2         $ & 5.75 &    & 6.15 & \\
CT Ser   & 0.1950 & 16.14 & 1.4  & $0.02\pm 0.17$ & 5.40 &    & 5.75 & \\
FH Ser   &        & 16.97 & 0.95 & $2.02\pm 0.50$ & 5.05 &    & 5.45 & \\
XX Tau   &        & 18.58 & 3.5  & $1.26\pm 0.57$ & 4.60 &    & 4.95 & \\
RW UMi   & 0.079  & 18.95 & 5    & $0.29\pm 0.24$ & 5.15 &    & 5.55 & \\
LV Vul   &        & 15.90 & 0.92 & $1.75\pm 0.42$ & 4.35 &    & 4.70 & \\
NQ Vul   &        & 17.70 & 1.16 & $2.1 \pm 0.7 $ & 5.30 &    & 5.65 & \\
PW Vul   & 0.2137 & 17.51 & 1.8  & $1.73\pm 0.32$ & 4.50 &    & 4.85 & \\
QU Vul   & 0.1118 & 18.31 & 1.75 & $1.89        $ & 5.20 &    & 5.55 & \\
QV Vul   &        & 18.61 & 2.7  & $1.26        $ & 5.20 &    & 5.55 & \\ \\
SS Aur   & 0.1828 & 10.5 & 0.20  & $0.19\pm 0.15$ & 3.80 & 38 & 4.40& Dwarf Novae\\
SS Cyg   & 0.2751 &  8.7 & 0.166 & $0.13\pm 0.17$ & 2.47 & 38 & 3.05& at maximum\\
U Gem    & 0.1729 &  9.8 & 0.096 & $0.27\pm 0.14$ & 4.62 & 70 & 3.90& \\ \\
AE Aqr   & 0.4117 & 11.5 & 0.102 & $0.20\pm 0.15$ & 6.26 & 58 & 6.20& Novalike\\
V3885 Sgr& 0.2163 & 10.3 & 0.11  & $0.08\pm 0.14$ & 5.01 &$<50$ & 5.40&Variables \\
RW Sex   & 0.2451 & 10.6 & 0.29  & $0.05\pm 0.17$ & 3.24 & 34 & 3.90& \\
IX Vel   & 0.1939 &  9.6 & 0.096 & $0.29\pm 0.21$ & 4.40 & 60 & 4.25& \\
\enddata

\end{deluxetable}


\begin{thebibliography}{}
\bibitem[Alm\'ar and Ill\'es-Alm\'ar(1966)]{alm66} Alm\'ar, I.,
         Ill\'es-Alm\'ar, E. 1966, Mitt. Sternw. Budapest 60
\bibitem[Andre\"a(1991)]{and91} Andre\"a, J. 1991, PhD thesis,
         Erlangen-N\"urnberg University
\bibitem[Andre\"a \etal(1991)]{ads91} Andre\"a, J., 
         Drechsel, H., Snijders, M.A.J., and Cassatella, A. 1991, \aap, 244, 111
\bibitem[Andre\"a, Drechsel, and Starrfield(1994)]{ads94} Andre\"a, J.,
         Drechsel, H., and Starrfield, S. 1994, \aap, 291, 869
\bibitem[Andrillat and Houziaux(1989)]{and89} Andrillat, Y. and Houziaux,
         L. 1989, \mnras, 238, 29p
\bibitem[Belle \etal(1999)]{belle99} Belle, K.E., Woodward, C. E.,
         Evans, A., Eyres, S., Gehrz, R. D., Schuster, M., Greenhouse, M. A.,
         Krautter, J., Starrfield, S. G., Truran, J. 1999, \baas, 31, 977
\bibitem[Bergner \etal(1985)]{ber85} Bergner, Y.K., Bondarenko, S.I., 
         Miroshnichenko, A.S., Yudin, R.V., Yutanov, N.Y., 
         Kuratov, K.S., and Mukanov, D.B. 1985, \astk, 1374
\bibitem[Bertaud(1945)]{ber45}Bertaud, C. 1945, Ann. Paris Obs. 9, fasc. 1
\bibitem[Bertaud(1962)]{ber62}Bertaud, C. 1962, \jo, 45, 321 
\bibitem[Bowen(1956)]{bow56} Bowen, I.S. 1956, \aj, 61, 336
\bibitem[Buscombe and de Vaucouleurs(1955)]{bdv55} Buscombe, W., de
         Vaucouleurs, G. 1955, \obs, 75, 170 
\bibitem[Campbell(1903)]{cam03} Campbell, L. 1903, \ha, 48, 39
\bibitem[Campbell(1919)]{cam19} Campbell, L. 1919, \ha, 81, 113
\bibitem[Ciatti and Rosino(1974)]{cia74} Ciatti, F. and Rosino, L. 1974, 
         \aaps, 16, 305
\bibitem[Chochol \etal(1997)]{cgp97} Chochol, D., Grygar, J., Pribulla, T., 
         Kom\v{z}\'\i k, R., Hric, L., and Elkin, V. 1997, \aap, 318, 908 
\bibitem[Chochol \etal(1993)]{chu93} Chochol, D., Hric, L., Urban, Z.,
         Kom\v{z}\'\i k, R., Grygar, J., and Papousek, J. 1993, \aap, 277, 103
\bibitem[Chochol \etal(1995)]{ckh95} Chochol, D., Kom\v{z}\'\i k, R., 
         Hric, L., and Grygar, J. 1995, in Cataclysmic Variables, 
         ed. A. Bianchini et al., (Dordrecht, Kluwer Academic Publishers), 
         p. 153
\bibitem[Cohen(1985)]{coh85} Cohen, J.G. 1985, \apj, 292, 90
\bibitem[Cohen and Rosenthal(1983)]{cr83} Cohen, J.G. and Rosenthal, A.J. 
         1983, \apj, 268, 689
\bibitem[Cottrell and Smith(1978)]{cs78}  Cottrell, M.J. and 
         Smith, S.E. 1978, \pasp, 90, 441
\bibitem[Della Valle(1991)]{del91} Della Valle, M. 1991, \aap, 252, L9
\bibitem[Della Valle(1993)]{del93} Della Valle, M. 1993, \iaucirc, 5870
\bibitem[Della Valle and Benetti(1996)]{dvb96} 
         Della Valle, M. and Benetti, S. 1996, \iaucirc, 6532
\bibitem[Della Valle and Duerbeck(1993)]{dd93} 
         Della Valle, M. and Duerbeck, H.W. 1993, \aap, 275, 239 
\bibitem[Della Valle and Livio(1995)]{dvl95} Della Valle, M. and Livio, 
         M. 1995, \apj, 452, 704
\bibitem[Della Valle \etal(1997)]{dgb97} 
         Della Valle, M., Gilmozzi, R., Bianchini, A., and Esenoglu, H. 
         1997, \aap, 325, 1151
\bibitem[Drechsel \etal(1977)]{drd77} Drechsel, H., Rahe, J., 
         Duerbeck, H.W., Kohoutek, L., and Seitter, W. 1977, \aaps, 30, 323
\bibitem[Duerbeck(1981)]{due81} Duerbeck, H.W. 1981, \pasp, 93, 165
\bibitem[Duerbeck(1987a)]{due87a} Duerbeck, H.W. 1987a, \apss, 131, 461
\bibitem[Duerbeck(1987b)]{due87b} Duerbeck, H.W. 1987b, \ssr, 45, 1
\bibitem[Duerbeck(1987c)]{due87c} Duerbeck, H.W. 1987c, The Messenger (ESO),
         50, 8
\bibitem[Duerbeck(1992)]{due92} Duerbeck, H.W. 1992, \aca, 42, 85
\bibitem[Duerbeck(1996)]{due96} Duerbeck, H.W. 1996, in Light Curves of 
         Variable Stars -- a Pictorial Atlas, ed. C. Sterken and C. Jaschek,
         (Cambridge, Cambridge University Press), p. 134 seq.
\bibitem[Duerbeck(1999)]{due99} Duerbeck, H.W. 1999, \ibvs, 4731
\bibitem[Duerbeck and Covarrubias(1995)]{due95} Duerbeck, H.W. and
         Covarrubias, R. 1995, in IAU Coll. 151, Flares and Flashes,
         eds. J. Greiner et al., (Berlin, Springer), p. 264
\bibitem[Duerbeck and Zwitter(2000)]{due00} Duerbeck, H.W. and Zwitter, T.
         2000, in preparation
\bibitem[Duerbeck \etal(1984)]{dgn84} Duerbeck, H.W., Geffert, M., Nelles, B.,
         D\"ummler, R., and Nolte, M. 1984, \ibvs, 2641
\bibitem[Duerbeck \etal(2000)]{dsls00} Duerbeck, H.W., Shara, M.M., Leibowitz,
         E.M., and Seitter, W.C. 2000, in preparation
\bibitem[Gaposchkin(1946)]{gap46} Gaposchkin, S. 1946, \hb, 918
\bibitem[Gehrz \etal(1985)]{geh85} Gehrz, R.D., Grasdalen, G.L., and
         Hackwell, J.A. 1985, \apjl, 298, L47
\bibitem[Gill and O'Brien(1998)]{go98} Gill, C.D. and O'Brien, T.J. 1998, 
         \mnras, 300, 221
\bibitem[Gill and O'Brien(2000)]{go00} Gill, C.D. and O'Brien, T.J. 2000, 
         \mnras, 314, 175
\bibitem[Gilmozzi, Selvelli and Cassatella(1994)]{gil94} Gilmozzi, R., 
         Selvelli, P., Cassatella, A. 1994, \memsai, 65, 199
\bibitem[Gratton(1953)]{gra53} Gratton, L. 1953, \apj, 118, 568
\bibitem[Haefner(1999)]{hae99} Haefner, R. 1999, \ibvs, 4706
\bibitem[Hakkila \etal(1997)]{hak97} Hakkila, J., Myers, J.M., Stidham, B.J.,
         Hartmann, D.H. 1997, \aj, 114, 2043
\bibitem[Herbig and Smak(1992)]{hs92}Herbig, G.H. and Smak, J.I. 1992, 
         \aca, 42, 17 
\bibitem[Hjellming(1994)]{hje94} Hjellming. R.M. 1994, talk presented at the
         184th AAS meeting
\bibitem[Kaluzny and Chlebowski(1989)]{kc89} Kaluzny, J. and Chlebowski, T.
         1989, \aca, 39, 35 
\bibitem[Kenyon and Wade(1986)]{kw86} Kenyon, S.J. and Wade, R.A. 1986, 
         \pasp, 98, 935
\bibitem[Klare and Wolf(1978)]{kw78} Klare, G. and 
         Wolf, B. 1978, \aaps, 33, 327 
\bibitem[Kohoutek(1981)]{koh81} Kohoutek, L. 1981, \mnras, 196, 87P
\bibitem[Krautter \etal(1981)]{kkw81} Krautter, J., Klare, G., Wolf, B.,
         Duerbeck, H.W., Rahe, J., Vogt, N., and Wargau, W. 1981, 
         \aap, 102, 337
\bibitem[Kukarkin and Gitz(1937)]{kuk37} Kukarkin, B.W. and Gitz, H.K. 1937,
         Astr. Zh. SU 14, 220
\bibitem[Lance, McCall and Uomoto (1988)]{lmu88} Lance, C.M., McCall, M.L.
         and Uomoto, A.K. 1988, \apjs, 66, 151
\bibitem[Larsson-Leander(1954)]{lar54} Larsson-Leander, G. 1954, An. Stockholm
         Obs. 18, No. 3
\bibitem[Leibowitz, Wyckoff and Vidal(1976)]{lwv76}  Leibowitz, E.M., 
         Wyckoff, S., and Vidal, N.V. 1976, \pasp, 88, 750
\bibitem[Liller(1993)]{lil93} Liller, W. 1993, \iaucirc, 5867
\bibitem[MacConnell(1999)]{mac99} MacConnell, D.J. 1999, private communication
\bibitem[Malakpur(1975)]{mal75} Malakpur, I. 1975, \apss, 48, 403
\bibitem[Marsh, Oke and Wade(1983)]{mar83}Marsh, T.R., Oke, J.B., and Wade,
         R.A. 1983, \mnras, 205, 33p 
\bibitem[Mattei(1999)]{mat99} Mattei, J. A., 1999, Observations from the 
         AAVSO International Database, private communication
\bibitem[McLaughlin(1943)]{mcl43} McLaughlin, D.B. 1943, Pub. AAS 10, 310
\bibitem[McLaughlin(1960)]{mcl60} McLaughlin, D.B. 1960, in Stellar
         Atmospheres, ed. J.L. Greenstein, (Chicago, University of Chicago
         Press), p. 585
\bibitem[Orio \etal(1996)]{obd96} Orio, M., Balman, S., Della Valle, M.,
         Gallagher, J. and \"Ogelman, H. 1996, \apj, 466, 410
\bibitem[Pachoulakis and Saizar(1995)]{ps95}  Pachoulakis, I. and Saizar,
         P. 1995, in Cataclysmic Variables, ed. A. Bianchini et al.,  
         (Dordrecht, Kluwer Academic Publishers), p. 153
\bibitem[Paresce(1994)]{par94} Paresce, F. 1994, \aap, 282, L13
\bibitem[Paresce \etal(1995)]{plhk95} Paresce, F., Livio, M., Hack, W., and 
         Korista, K. 1995, \aap, 299, 823
\bibitem[Pettit(1949)]{pet49} Pettit, E. 1949, \pasp, 61,41
\bibitem[Ringwald and Naylor(1996)]{rn96} Ringwald, F.A. and Naylor, T. 
         1996, \mnras, 278, 808
\bibitem[Ringwald, Naylor and Mukai (1996)]{rnm96} Ringwald, F.A. 
         Naylor, T. and Mukai. K. 1996, \mnras, 281, 192
\bibitem[Ringwald \etal(1998)]{rwo98} Ringwald, F.A., 
         Wade, R.A., Orosz, J.A., and Ciardullo, R.B.
         1998, \baas, 30, 893
\bibitem[Ritter and Kolb(1998)]{rit98} Ritter, H. and Kolb, U. 1998, \aaps,
         129, 83 
\bibitem[Robb and Scarfe(1995)]{rs95} Robb, R.M. and Scarfe, C.D. 1995, \mnras,
         273, 347 
\bibitem[Robertson, Honeycutt and Henden(1999)]{rhh99} Robertson, J.W., 
         Honeycutt, R.K., and Henden, A.H. 1999, \baas, 31, 1247 
\bibitem[Rosino and Iijima(1987)]{ri87} Rosino, L. and Iijima, T. 1987, 
         \apss, 130, 157
\bibitem[Rosino \etal(1996)]{ros96} Rosino, L., Iijima,  T., Rafanelli, P., 
         Radovich, M., Esenoglu, H. and Della Valle 1996, \aap, 315, 463
\bibitem[Saizar \etal(1991)]{sai91} Saizar, P., Starrfield, S., Ferland, G. J.,
         Wagner, R. M., Truran, J. W., Kenyon, S. J., Sparks, W. M., 
         Williams, R. E., Stryker, L. L. 1991, \apj, 367, 310
\bibitem[Saizar \etal(1996)]{sps96} Saizar, P., Pachoulakis, I., Shore, S.N.,
         Starrfield, S., Williams, R.E., Rothschild, E. and Sonneborn, G., 
         1996, \mnras, 279, 280
\bibitem[Sanford(1945)]{san45} Sanford, R.F. 1945, \apj, 102, 357
\bibitem[Satyvaldiev(1963)]{sat63} Satyvaldiev, V. 1963,
         Byull.~Inst.~Astrof.~AN~Tadj.~SSR (Dushanbe) 36, 37
\bibitem[Seitter(1990)]{sei90} Seitter, W.C. 1990, in IAU Coll. No. 122,
         Physics of Classical Novae, ed. A. Cassatella and R. Viotti, 
         (Berlin, Springer-Verlag), p. 79
\bibitem[Sekiguchi \etal(1989)]{sek89} Sekiguchi, K. Feast, M.W., 
         Fairall, A.P., Winkler, H. 1989, \mnras, 241, 311
\bibitem[Shin \etal(1998)]{shi98} Shin, J.-Y., Gehrz, R.D., Jones, T.J., 
         Krautter, J., Heidt, J., and Hjellming, R.M. 1998, \aj, 116, 1966
\bibitem[Shore \etal(1993)]{sho93} Shore, S.N., Sonneborn, G., 
         Starrfield, S., Gonzalez-Riestra, R. and Ake, T.B. 1993, 
         \aj, 106, 2408
\bibitem[Slavin, O'Brien and Dunlop(1994)]{sod94} Slavin, A.J., O'Brien, T.J.
         and Dunlop, J.S. 1994, \mnras, 266, L55
\bibitem[Slavin, O'Brien and Dunlop(1995)]{sod95} Slavin, A.J., 
         O'Brien, T.J. and Dunlop, J.S. 1995 \mnras, 276, 353
\bibitem[Soderblom(1976)]{sod76} Soderblom, D. 1976, \pasp, 88, 517
\bibitem[Solf(1983)]{solf83} Solf, J. 1983, \apj, 273, 647
\bibitem[Spencer Jones(1931)]{spe31} Spencer Jones, H. 1931, Ann. Cape Obs. 10
\bibitem[Szkody(1994)]{szk94} Szkody, P. 1994, \aj, 108, 639
\bibitem[Szkody and Howell(1992)]{sh92} Szkody, P. and Howell, S.B. 1992, 
         \apjs, 78, 537
\bibitem[Taylor \etal(1987)]{tshp87} Taylor, A.R., Seaquist, E.R., 
         Hollis, J.M., and Pottasch, S.R. 1987, \aap, 183, 38
\bibitem[Thorstensen and Taylor(2000)]{tho00} Thorstensen, J.R., and Taylor,
         C.J. 2000, \mnras, 312, 629
\bibitem[van den Bergh and Pritchet(1986)]{vbp86} van den Bergh, S. and
         Pritchet, C.J. 1986, \pasp, 98, 110  
\bibitem[van den Bergh and Younger(1987)]{vdB87} van den Bergh, S. and Younger
         P.F. 1987, \aaps, 70, 125 
\bibitem[Verbunt(1987)]{ver87}Verbunt, F. 1987, \aaps, 71, 339
\bibitem[Vogt and Maitzen(1977)]{vm77} Vogt, N. and Maitzen, H.M. 1977, 
         \aap, 61, 601
\bibitem[Wade \etal(1991)]{wcj91} Wade, R.A., Ciardullo, R., Jacoby, G.H.,
         and Sharp, N.A. 1991, \aj, 102, 1738
\bibitem[Wade \etal(2000)]{whc00} Wade, R.A., Harlow, J.J.B., and Ciardullo, R.
         2000, \pasp, 112, 614
\bibitem[Warner(1987)]{war87} Warner, B. 1987, \mnras, 227,83
\bibitem[Warner(1995)]{war95} Warner, B. 1995, Cataclysmic Variable Stars,
(Cambridge, Cambridge University Press), p. 146
\bibitem[Williams(1982)]{wil82} Williams, R.E. 1982, \apj, 261, 170
\bibitem[Williams(1992)]{wil92} Williams, R.E. 1992, \aj, 104, 725
\bibitem[Williams, Phillips, and Hamuy(1994)]{wil94} Williams, R.E., 
         Phillips, M.M., and Hamuy, M. 1994, \apjs, 90, 297
\bibitem[Whitney and Clayton(1989)]{whi89} Whitney, B.A. and Clayton, G.C.
         1989, \aj, 98, 297
\bibitem[Wu \etal(1989)]{wph89} Wu, C.-C., Panek, R.J., Holm, A.V., Raymond,
         J.C., Hartmann, L.W., and Swank, J.H. 1989, \apj, 339, 443
\bibitem[Younger(1980)]{you80} Younger, J.W. 1980, \aj, 85, 1233  
\bibitem[Zakharov(1953)]{zak53} Zakharov, G.P. 1953, \perz, 9, 175

\end{thebibliography}
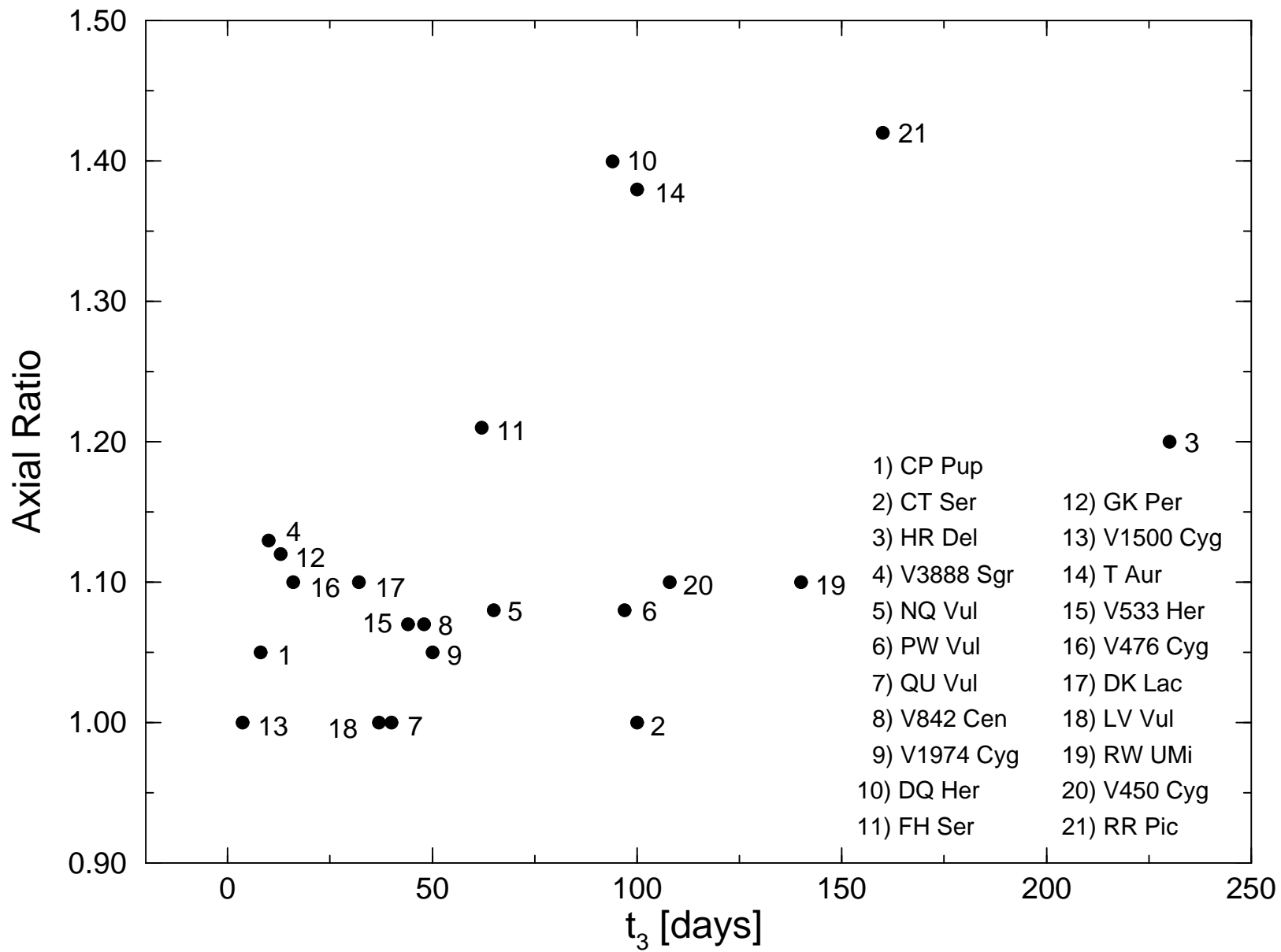
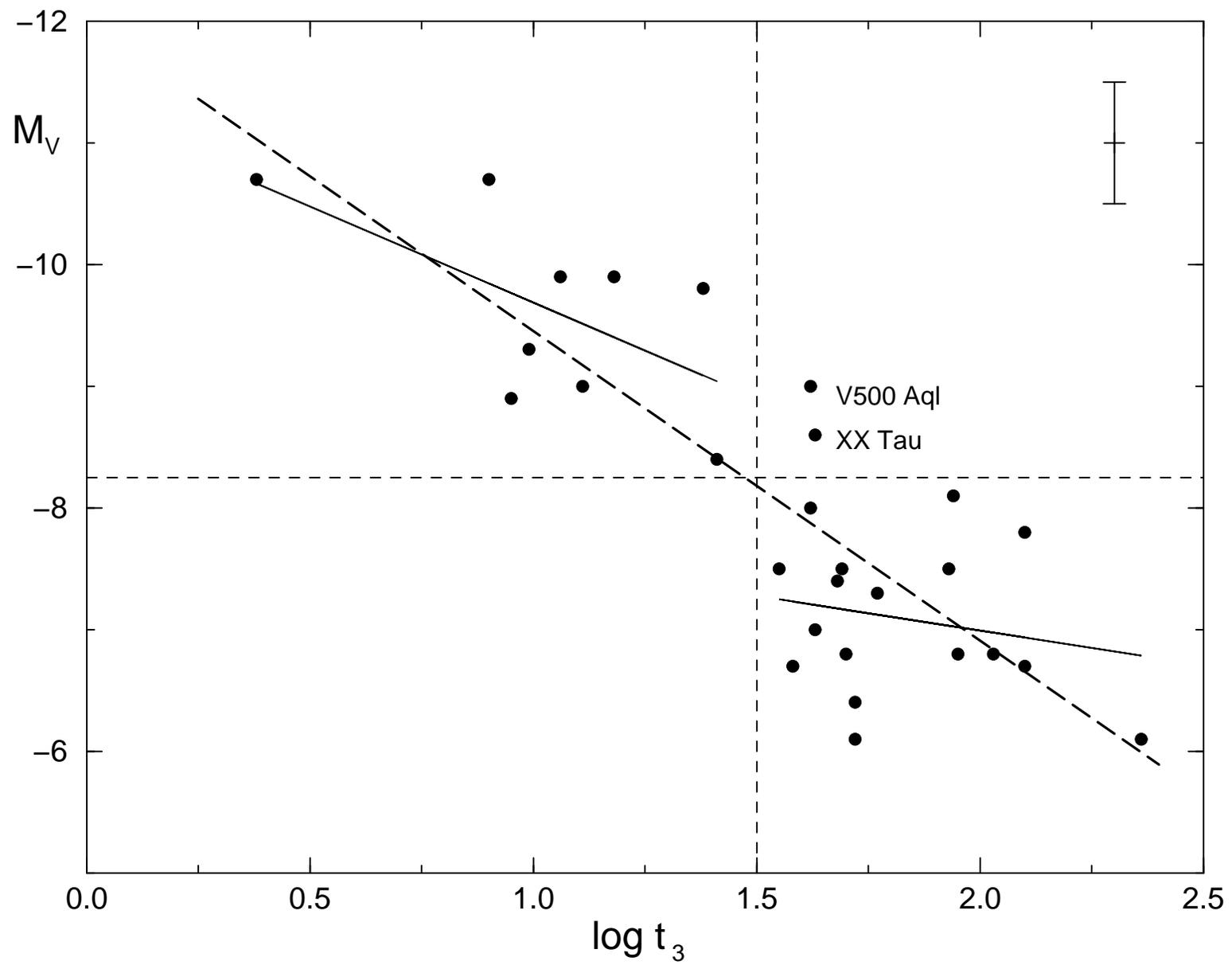
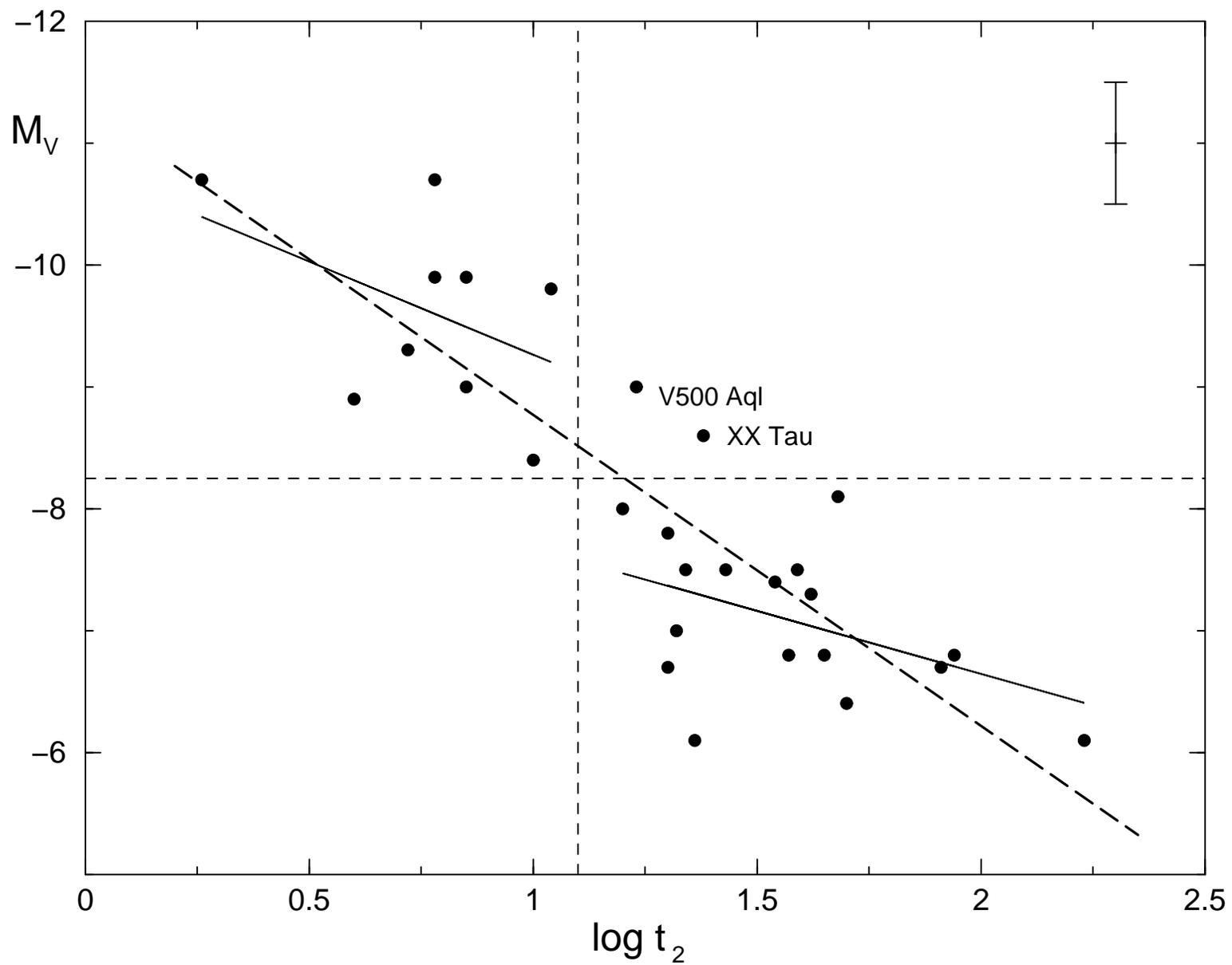
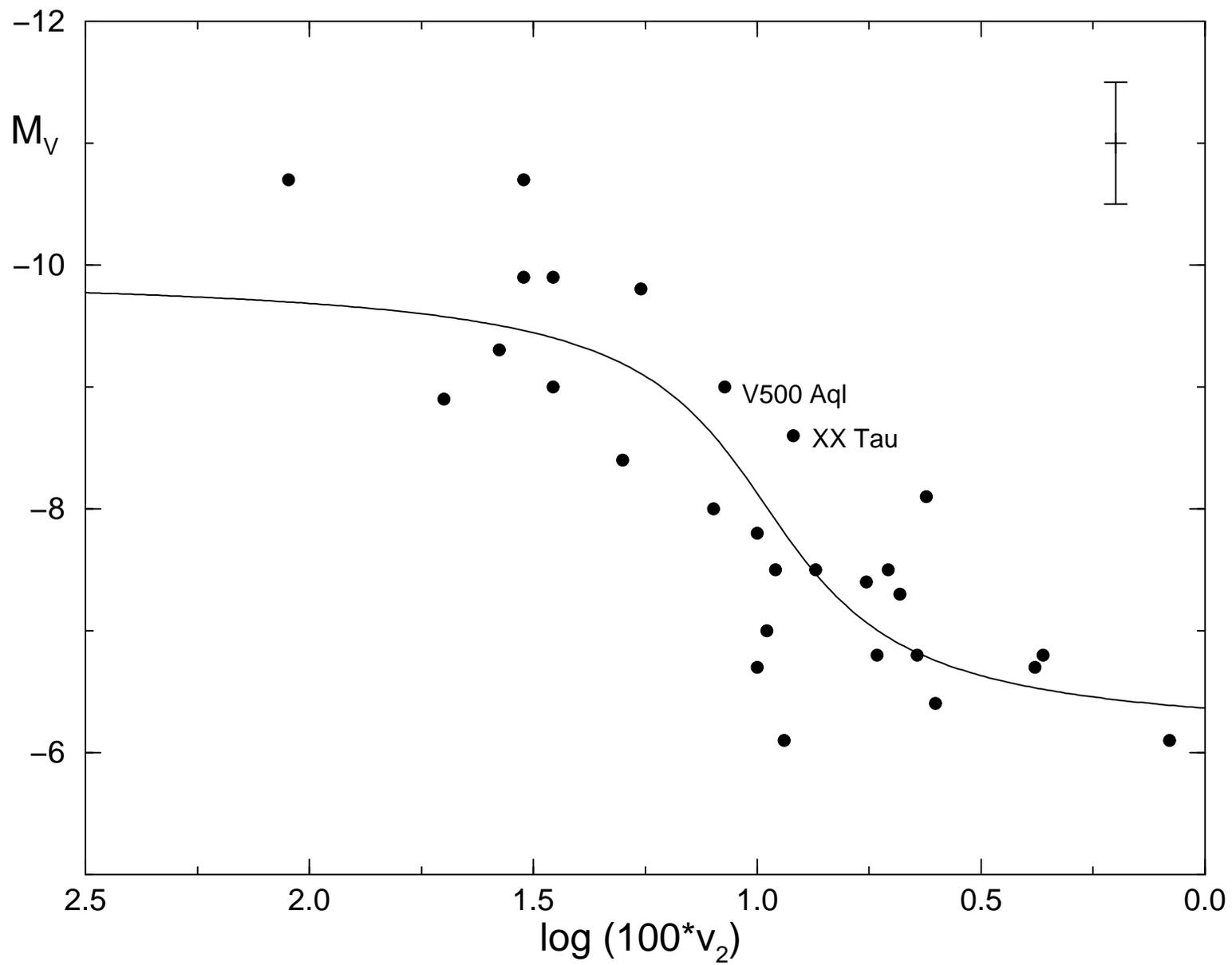
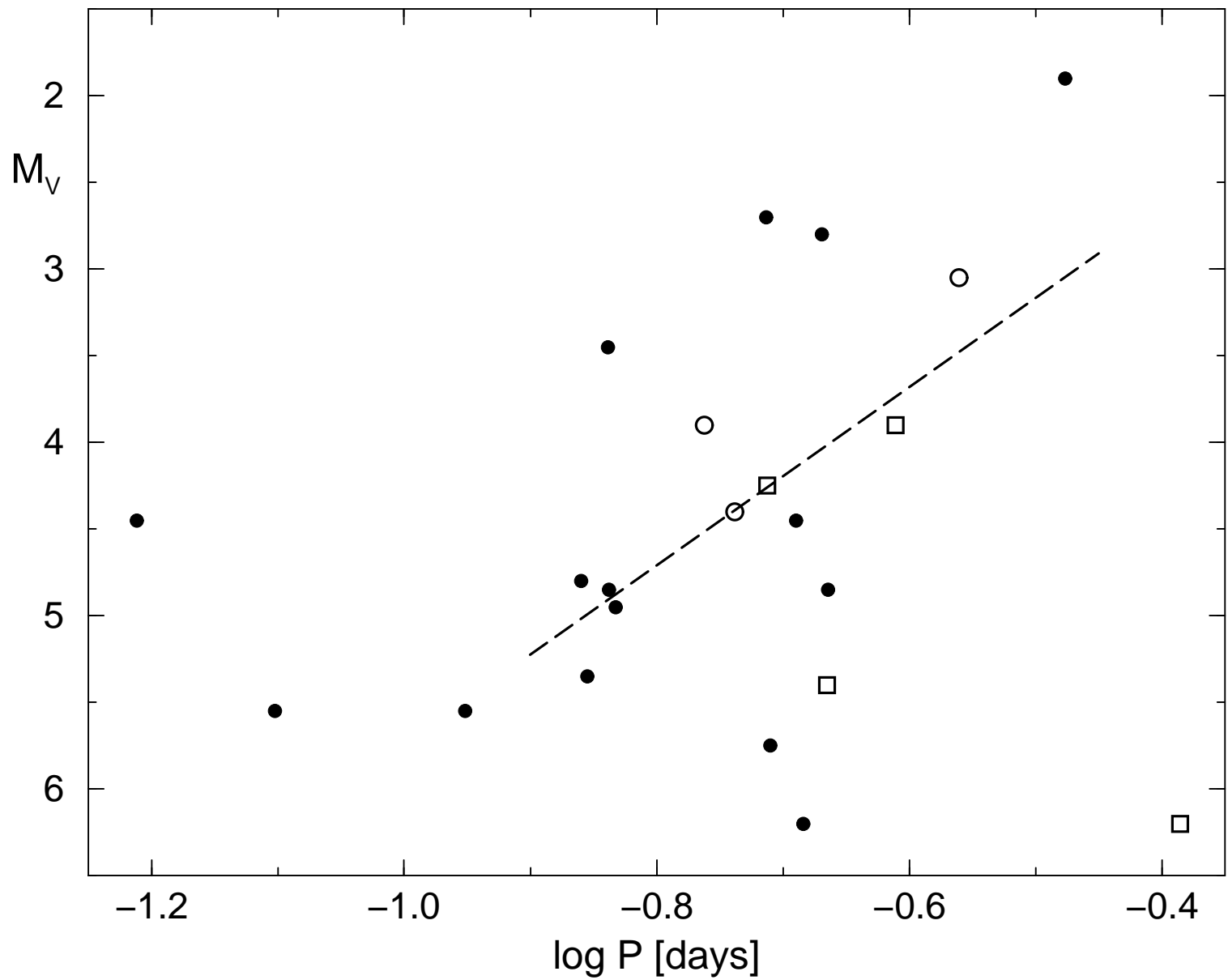
\end{document}